\documentclass[]{pasj02} 
\usepackage[switch,mathlines]{lineno} 
\usepackage{scalefnt}
\usepackage{natbib}
\usepackage{url}
\usepackage{ulem}
\usepackage{amsmath}
\usepackage{rotating}

\jyear{2026}
\Received{2026/06/22}
\Accepted{2026/08/12}

\begin{document} 

\title{First Trigonometric Parallax Measurements with the KVN and VERA Array (KaVA)}

\author{
 Nobuyuki \textsc{Sakai}\altaffilmark{1, 2}\altemailmark\orcid{0000-0002-5814-0554} \email{nobuyuki@narit.or.th; nobuyuki.sakai@nao.ac.jp}, Daisuke \textsc{Sakai}\altaffilmark{2, 3},
  Noriyuki \textsc{Kawaguchi}\altaffilmark{2}\orcid{0000-0002-7776-3159}, 
Shuangjing \textsc{Xu}\altaffilmark{4, 5}\orcid{0000-0003-2953-6442}, 
Bo \textsc{Zhang}\altaffilmark{5}\orcid{0000-0003-1353-9040}, 
Taehyun \textsc{Jung}\altaffilmark{4}\orcid{0000-0001-7003-8643}, 
Chungsik \textsc{Oh}\altaffilmark{4}, 
Jeong-Sook \textsc{Kim}\altaffilmark{2}\orcid{0000-0002-4987-5540}, 
Soon-Wook \textsc{Kim}\altaffilmark{4}\orcid{0000-0002-0978-4775}, 
Hiroshi \textsc{Imai}\altaffilmark{7}\orcid{0000-0002-0880-0091},
Yuanwei \textsc{Wu}\altaffilmark{8},
Maria \textsc{Rioja}\altaffilmark{9}\orcid{0000-0003-4871-9535},
Niu \textsc{Liu}\altaffilmark{10},
Zehao \textsc{Lin}\altaffilmark{11},
Ibnu Nurul \textsc{Huda}\altaffilmark{12},
Shuaibo \textsc{Bian}\altaffilmark{11}\orcid{0000-0002-7508-9615},
Leonid \textsc{Petrov}\altaffilmark{13}\orcid{0000-0001-9737-9667},
Jingdong \textsc{Zhang}\altaffilmark{14}\orcid{0000-0002-9768-2700},
Takaaki \textsc{Jike}\altaffilmark{2}, 
Yoshiaki \textsc{Tamura}\altaffilmark{15}, 
Jeong Ae Lee\altaffilmark{4}, 
\textsc{Hao Ding}\altaffilmark{6}\orcid{0000-0002-9174-638X}, 
Koichiro \textsc{Sugiyama}\altaffilmark{1}\orcid{0000-0002-6033-5000}, 
Kazuhiro \textsc{Hada}\altaffilmark{16}\orcid{0000-0001-6906-772X}, 
Kiyoaki \textsc{Wajima}\altaffilmark{4}\orcid{0000-0003-3823-7954}, 
Kazuya Hachisuka\altaffilmark{2}, 
Mareki Honma\altaffilmark{2}\orcid{0000-0003-4058-9000}, 
Kee-Tae Kim\altaffilmark{4, 17}\orcid{0000-0003-2412-7092},
and Se-Jin Oh\altaffilmark{4}
}

 \altaffiltext{1}{National Astronomical Research Institute of Thailand, 260 Moo 4, T.Donkaew, A. Maerim, Chiang Mai, 50180, Thailand}
 

\altaffiltext{2}{Mizusawa VLBI Observatory, National Astronomical
  Observatory of Japan, 2-12 Hoshigaoka, Mizusawa, Oshu, Iwate 023-0861, Japan}
  

\altaffiltext{3}{Tokyo Electron Ltd., 52 Esashi-iwayado Matsunagane, Oshu, Iwate 023-1101, Japan}

\altaffiltext{4}{Korea Astronomy $\&$ Space Science Institute, 776, Daedeokdae-ro, Yuseong-gu, Daejeon 34055, Korea}

\altaffiltext{5}{Shanghai Astronomical Observatory, Chinese Academy of Sciences 80 Nandan Road, Shanghai, 200030, China}


\altaffiltext{6}{National Astronomical Observatories, Chinese Academy of Sciences, Beijing 100012, China}


 \altaffiltext{7}{Amanogawa Galaxy Astronomy Research Center, Graduate School of Science and Engineering, Kagoshima University, 1-21-35 Korimoto Kagoshima}

 \altaffiltext{8}{National Time Service Center, Chinese Academy of Sciences, Xi’an 710600, China}

\altaffiltext{9}{Int. Centre for Radio Astronomy Research, University of Western Australia, Perth, Australia}

\altaffiltext{10}{School of Astronomy and Space Science, Key Laboratory of Modern Astronomy and Astrophysics (Ministry of Education),
Nanjing University, Nanjing 210023, China}

\altaffiltext{11}{Purple Mountain Observatory, Chinese Academy of Sciences, Nanjing 210023, China}

\altaffiltext{12}{Research Center for Computing, National Research and Innovation Agency, Bogor, Indonesia}

\altaffiltext{13}{NASA Goddard Space Flight Center, Code 61A, 8800 Greenbelt Rd, Greenbelt, 20771 MD, USA}


\altaffiltext{14}{Department of Geodesy and Geodynamics, Finnish Geospatial Research Institute (FGI), National Land Survey of Finland, Vuorimiehentie 5, Espoo, FI-02150, Finland}

\altaffiltext{15}{Iwate Industrial Technology Junior College, 66-2 Higashihiromachi Sakurakawa, Mizusawa, Oshu, Iwate 023-0003, Japan}

\altaffiltext{16}{Graduate School of Science, Nagoya City University,
Yamanohata 1, Mizuho-cho, Mizuho-ku, Nagoya, 467-8501, Aichi,
Japan}




\altaffiltext{17}{University of Science and Technology, Korea (UST), 217 Gajeong-ro, Yuseong-gu, Daejeon 34113, Republic of Korea}




\maketitle

\begin{abstract}
To demonstrate the astrometric capability of the combined Korean VLBI (Very Long Baseline Interferometry) Network (KVN) and VLBI Exploration of Radio Astrometry (VERA) Array (KaVA), we conducted six-epoch VLBI observations of 22-GHz H$_{2}$O masers associated with the star-forming region W3(OH). Two atmospheric calibration methods, (1) GPS and (2) JMA (Japan Meteorological Agency) mesoscale analysis data, and two phase-reference sources were independently applied to the astrometric analysis. Trigonometric parallaxes of W3(OH) were successfully measured with all calibration strategies, including the first successful parallax measurement using JMA calibration. The combined-fit parallax is 0.497$\pm$0.024 mas, corresponding to a distance of 2.01$^{+0.10}_{-0.09}$ kpc. This value is consistent with the previous Very Long Baseline Array (VLBA) result of 0.489$\pm$0.017 mas. KaVA achieved a 55$\%$ higher signal-to-noise ratio in phase-referenced maps than VERA, in good agreement with theoretical expectations. These results suggest that KaVA will enable trigonometric parallax measurements of 22-GHz H$_2$O masers that are difficult to observe with VERA alone because of low flux densities and/or limited $uv$ coverage. A flux variation from 510\,$\pm$\,50 to 3900\,$\pm$\,400 Jy was detected in W3(OH) over a one-year observing campaign and is attributed to two nearby maser features separated by only $\sim$64 AU at 2.01 kpc. The brightest feature showed a decrease in linewidth with increasing peak flux density, consistent with unsaturated maser amplification, whereas no similar trend was found for the other feature, suggesting different responses to the same amplification conditions, possibly owing to differences in saturation state and internal velocity structure.


\end{abstract}

\pagewiselinenumbers 

\section{Introduction}
The East Asian VLBI Network (EAVN) is rapidly expanding (e.g., \citealp{2022Galax..10..113A}). The Korean VLBI Network (KVN) and the Japanese VERA (VLBI Exploration of Radio Astrometry) have been successfully combined to form the KVN and VERA Array (KaVA), which operates at 22 and 43 GHz (\citealp{2013arXiv1310.2705S,2014PASJ...66..103N,2014ApJ...789L...1M}). Regular open-use observations with KaVA have been conducted since 2014. As of 2026, the EAVN open-use program, comprising 17 radio telescopes (KaVA = 8 telescopes, Tianma 65m, Sheshan 25m, Nanshan 26m, Takahagi 32m, Hitachi 32m, Yamaguchi 32m, Sejong 22m, Nobeyama 45m, and Kunming 40m), is available to the international community. Future collaborations with radio telescopes in Southeast Asia are also expected (e.g., \citealp{2022arXiv221004926J}; \citealp{2024evn..conf..177S}).
While most scientific results obtained with KaVA/EAVN have been based on imaging observations (e.g., \citealp{2014ApJ...789L...1M,2015PKAS...30..453K,2016PASJ...68...77A,2017PASJ...69...71H,2018evn..confE..75B,2019MNRAS.486.2412L,2020ApJ...901....2H,2021RAA....21..205C,2022ApJ...932...64B,2023Natur.621..711C,2024MNRAS.52710031R,2026A&A...707A.191T}), the astrometric capability of KaVA/EAVN, particularly for parallax measurements, remains to be fully evaluated (e.g., \citealp{2020SciBu..65..267A}). Proper motion measurements with KaVA/EAVN have been reported in a limited number of studies (e.g., \citealp{2021PASJ...73.1669T,2023ApJ...943...79A,2023PASJ...75..208S,2025PASJ...77..678K,2026ApJ..1000L..45J}). In addition, several KaVA results have been obtained using the frequency phase-referencing technique (e.g., \citealp{2016ApJ...822....3Y,2019JKAS...52...23Z}). Although \citet{2022ApJ...941..105X} studied the astrometric evolution of H$_{2}$O masers associated with BX Cam using the EAVN, the astrometric analysis was based solely on VERA data. Therefore, the parallax measurement capability of KaVA and EAVN needs to be evaluated and demonstrated to broaden the range of possible science cases, as discussed by \citet{2014PASJ...66..105D}.

To demonstrate the astrometric capability of KaVA, the star-forming region W3(OH) is an ideal, as independent VLBI parallax measurements have been reported for this source, allowing a direct comparison with previous results.
\citet{2006ApJ...645..337H} measured a trigonometric parallax of 0.489$\pm$0.017 mas at 22 GHz toward W3(OH) using the VLBA, while \citet{2006Sci...311...54X} reported a value of 0.512$\pm$0.010 mas at 12 GHz for the same region, also using the VLBA. In addition, \citet{2011PASJ...63.1345M} measured a trigonometric parallax of 0.598$\pm$0.067 mas at 6.7 GHz toward W3(OH) using VERA. The coordinates adopted in previous VLBI parallax studies are consistent within $\sim$0.03" between the 12 and 6.7 GHz CH$_3$OH masers, whereas the 22 GHz H$_2$O masers are located approximately 7" away from the methanol maser position. For sources other than W3(OH), VLBI parallaxes have also been measured multiple times for the same targets using different VLBI arrays, observing frequencies, and data-analysis methods (e.g., \citealp{2010A&A...511A...2R, 2014ApJ...787...54A, 2019ApJ...885..131R}). Previous studies have reported cases in which the resulting parallaxes were consistent, as well as cases in which they were not. Therefore, even for sources with existing VLBI parallax measurements, follow-up observations under conditions different from those of previous studies are important. 



\citet{1993LNP...412..225A} reported that the flux density of individual 22-GHz H$_2$O maser components in the W3(OH) region can undergo significant variations on timescales of months. In \citet{2006ApJ...645..337H}, the flux density of 22-GHz H$_2$O maser components was found to vary by factors of $\sim$2–30 during the observations. Under unsaturated conditions, where the population inversion responsible for maser emission is governed solely by pumping rather than by the radiation field, \citet{1992ASSL..170.....E} showed that the maser intensity $I_\nu$ responds exponentially to variations in the pumping conditions, as
\begin{equation}
\label{Eq:maser-unsaturated}
I_\nu \propto \mathrm{exp}(\kappa_0 \ell) = \mathrm{exp}(\tau_0)
\end{equation}
where $\kappa_0$ is the absorption (or gain) coefficient at the line center, $\ell$ is the effective path length contributing to the amplification, and $\tau_0$ is the optical depth at the line center, defined as a positive quantity for amplification. Changes in the local gas density and kinetic temperature can modify the gain coefficient at the line center, $\kappa_0$ (e.g., \citealp{1996ApJ...456..250K}), while velocity coherence along the line of sight and geometrical effects can alter the effective path length, $\ell$. In contrast, in the saturated regime, the maser intensity increases linearly with path length as $I_\nu \propto \ell$. Spectral information, such as linewidth and line-profile variations, together with VLBI imaging of the spatial distribution, enables us to investigate how and where the maser flux variability occurs.

VLBI astrometric accuracy is limited by delay-calibration errors (e.g., \citealp{2014ARA&A..52..339R}; \citealp{2020A&ARv..28....6R}). The tropospheric delay error is dominant at frequencies ($\nu$) higher than $\sim$10 GHz, while the ionospheric delay error dominates at $\nu$ $<$ 10 GHz. GPS (Global Positioning System) data and mini-geodetic observations (i.e., geodetic blocks) have been widely used for tropospheric calibration in VLBI astrometry, and more than 200 parallaxes have been measured for Galactic star-forming regions and AGB stars (e.g., \citealp{2019ApJ...885..131R}; \citealp{2020PASJ...72...50V}). \citet{2015PASJ...67...65N} compared tropospheric zenith delays derived from GPS observations with those estimated from the Japan Meteorological Agency (JMA) meso-scale analysis (MA) data. They concluded that both methods can calibrate the tropospheric zenith delay with an accuracy of $\sim$2 cm in terms of $c$$\tau_{{\rm err}}$, where $c$ is the speed of light and $\tau_{{err}}$ is the delay calibration error. However, trigonometric parallax measurement based on JMA calibration have not yet been demonstrated. Demonstrating consistent astrometric results using multiple independent tropospheric calibration methods is important for evaluating the robustness of VLBI astrometry. In addition, JMA-based calibration has practical advantages because it does not require GPS instrumentation at each station. It can also provide an alternative calibration strategy when GPS data are unavailable due to instrumental or operational problems.

In this paper, we evaluate the astrometric capability of KaVA at 22 GHz by measuring the trigonometric parallax of W3(OH). Our primary objective is to assess the agreement between the parallax obtained with KaVA and that previously measured with the VLBA, rather than to improve the precision of the previously reported parallax measurement of W3(OH). The consistency between the two measurements provides an initial assessment of the differential astrometric accuracy achievable with KaVA. Observations and data reduction are described in Sections 2 and 3, respectively. In Section 4, we present the results of the parallax and proper-motion measurements obtained using different tropospheric calibration methods and different phase-reference sources. In Section 5, we discuss these results, along with the flux variability of W3(OH). Finally, we summarize the paper in Section 6.   

\begin{table*}[htbp] 
\caption{Observation summary. \hspace{10em}} 
\begin{center} 
\label{table:1}
\small 
\begin{tabular}{lllcclc} 
\hline 
\hline 
Target&Right ascension &Declination 	  &$\Delta\nu$&$\Delta B_{\rm{IF}}$ &\multicolumn{1}{c}{Observation date} &Participating antennas 	  \\
	       && &(kHz)&(MHz)&\multicolumn{1}{c}{(UTC)}&\\ 
\hline
W3(OH)	&$\timeform{02h27m04s.8362}$  	&+$\timeform{61D52'24''.607}$	&31.25			&16 & 15:10--23:05 on 2017 Sep 5&12-4567 \\              
 	&  	&&	&& 08:25--16:35 on 2017 Dec 18 &1234567 \\

&& & && 04:48--12:58 on 2018 Feb 11    &12345\sout{6}\footnotemark[$*$]7\\  
 	&  	& 	&&& 20:00--28:09 on 2018 June 6&1234567 \\ 

&&&&&15:09--23:19 on 2018 Sep 3&1234-\sout{6}\footnotemark[$\dag$]7  \\
&&&&&15:09--23:19 on 2018 Sep 17&1234\sout{567}\footnotemark[$\ddag$] \\
\hline 
\multicolumn{4}{@{}l@{}}{\hbox to 0pt{\parbox{165mm}{\normalsize
\par\noindent
\\
Column 1: 22 GHz H$_{2}$O maser; Columns 2-3: equatorial coordinates (J2000.0); Column 4: frequency spacing; Column 5: IF bandwidth; Column 6: observation date and time (UTC); Column 7: participating antennas (1 = Mizusawa; 2 = Iriki; 3 = Ogasawara; 4 = Ishigaki-jima; 5 = Yonsei; 6 = Ulsan; 7 = Tamna). 

\footnotemark[$*$] Incorrect frequency setup at KVN Ulsan.\\
\footnotemark[$\dag$] Data recording problem at KVN Ulsan.\\
\footnotemark[$\ddag$] Incorrect polarization setup at all KVN stations.
}\hss}}
\end{tabular} 
\end{center} 
\end{table*}


\section{Observations}
Between 2017 September 5 and 2018 September 17, six VLBI astrometric observations were carried out with KaVA toward the star-forming region W3(OH) (see Table \ref{table:1}). We observed the H$_{2}$O maser emission from W3(OH) at a rest frequency of 22.235080 GHz. For relative VLBI astrometry, the target source W3(OH) was observed together with two phase-reference sources, J0223+6307 (ICRF J022329.6+630717) and J0244+6228 (ICRF J024457.6+622806). VERA employed its dual-beam system, in which the A-beam was pointed toward the target source while the B-beam was alternately switched between the two phase-reference sources. In contrast, KVN, using a single-beam system, alternately observed the target source and each phase-reference source. The fast-switching cycle time between W3(OH) and a phase-reference source was 60 seconds, and each W3(OH)-reference source pair was continuously observed for approximately 15 minutes before switching to the other phase-reference source pair for the next $\sim$15 minutes. According to the Radio Fundamental Catalogue (e.g., \citealp{2025ApJS..276...38P}), the coordinates adopted for J0223+6307 and J0244+6228 were $(\alpha, \delta)_{\rm J2000.0}$ = ($\timeform{02h23m29s.6070}$, +$\timeform{63D07'17''.306}$) and ($\timeform{02h44m57s.6967}$, +$\timeform{62D28'06''.516}$), respectively. The separation angle (SA) and position angle (PA) of J0223+6307 relative to W3(OH) are 1.3 deg and $-$18 deg, respectively. Note that PA is measured from north through east (i.e., PA = 90$^\circ$ corresponds to east). The SA and PA of J0244+6228 relative to the target are 2.2 deg and 72 deg, respectively. The peak flux density of W3(OH) varied from 510\,$\pm\,3\,(\mathrm{stat})\pm\,50\,(\mathrm{sys})$ to 3900\,$\pm\,1\,(\mathrm{stat})\pm\,400\,(\mathrm{sys})$ Jy during the observations. The flux density of J0223+6307 ranged from 21\,$\pm\,3\,(\mathrm{stat})\pm\,2\,(\mathrm{sys})$ to 130\,$\pm\,2\,(\mathrm{stat})\pm\,10\,(\mathrm{sys})$ mJy, while that of J0244+6228 ranged from 440\,$\pm\,6\,(\mathrm{stat})\pm\,40\,(\mathrm{sys})$ to 1000\,$\pm\,6\,(\mathrm{stat})\pm\,100\,(\mathrm{sys})$ mJy. The statistical and systematic errors are denoted in the above flux densities.

Each observing epoch had a total duration of about 8 hours and consisted of (i) four 30-minute ``geodetic blocks'', spaced approximately 2 hours apart for clock and atmospheric (tropospheric) delay calibration; (ii) four 5-minute ``manual phase-calibration'' scans of the continuum source DA55 every $\sim$2 hours; and (iii) fast-switching scans between the maser source W3(OH) and two phase-reference sources, J0223+6307 and J0244+6228. In each geodetic block, 16 bright continuum sources (International Celestial Reference Frame sources, hereafter ICRF sources) were observed with an integration time of 1 minute per source, covering a wide range of azimuth and elevation angles at each telescope. Left-handed circular polarization data were recorded at each telescope at a data rate of 1024 Mbps with 2-bit quantization, and were correlated with the KJCC/Daejeon FX-type hardware correlator\footnote{\url{https://radio.kasi.re.kr/kjcc/main.php}} \citep{Oh2010KJJVC,2015JKAS...48..125L}. An integration time of 1.6384 s was used for the correlation. Details of the VERA and KVN back-end systems are summarized in the EAVN Status Report \footnote{\url{https://radio.kasi.re.kr/status_report.php?cate=EAVN}}. The maser (line) data consisted of a single Intermediate Frequency (IF) band with a bandwidth of 16 MHz and were correlated into 512 spectral channels, yielding a frequency spacing of 31.25 kHz, corresponding to a velocity spacing of 0.42 km s$^{-1}$ at a rest frequency of 22.235080 GHz. In contrast, the continuum data consisted of 15 consecutive IF bands, each with a bandwidth of 16 MHz, resulting in a total bandwidth of 240 MHz. The starting frequency of IF 9 is identical to that of the line data. Each IF band was correlated into 64 spectral channels.

\section{Data reduction}


Data reduction was performed using the NRAO Astronomical Image Processing System ($AIPS$; \citealt{1996ASPC..101...37V}). We followed the procedures described in \citet{2023PASJ...75..208S} and \citet{2020PASJ...72...52N}. For time ranges during which KVN was not observing either the target or one of the phase-reference sources due to antenna fast switching, an $\texttt{FG}$ table was generated in AIPS and applied to the KVN data. The spatial distribution of H$_{2}$O maser spots was obtained for each dataset using phase referencing. We selected long-lived (i.e., $>$ 1 year) maser spots for the parallax and proper motion measurements. The data analysis procedures for amplitude and phase calibration are described below.

\subsection{Amplitude calibration}
The chopper-wheel (i.e., R/Sky) method (\citealp{1973ARA&A..11...51P}; \citealp{1976ApJS...30..247U}) was applied for the amplitude calibration of the KaVA data. The typical accuracy of the absolute amplitude calibration achieved with this method is approximately 10\,$\%$ \citep{1976ApJS...30..247U}. In this method, the visibility amplitudes were scaled in units of Jansky using the a priori antenna gains and the measured system noise temperatures corrected for atmospheric absorption (i.e., $T_{\rm sys}^{*}$) at each telescope. 

Following the recommendation of \citet{2015JKAS...48..229L}, a correction factor of 1.3 was applied to the visibility amplitudes of the KJCC data using the AIPS task $\texttt{APCAL}$.
The residual amplitude gains were further corrected through iterative self-calibration using the AIPS tasks $\texttt{CALIB}$ and $\texttt{IMAGR}$ for the individual phase-reference sources.

\begin{figure}[tbhp] 
 \begin{center} 
     \includegraphics[scale=1.0]{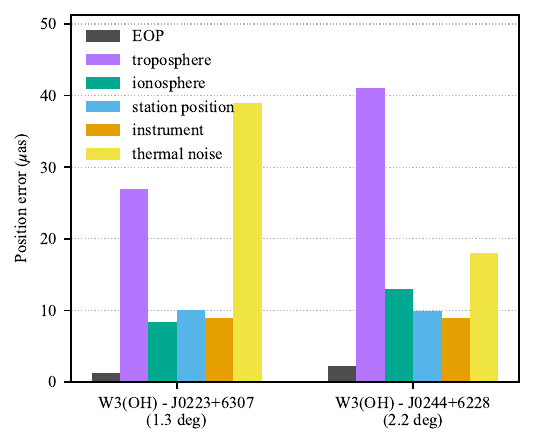} 
\end{center} 
\caption{ Expected position error budget for the VLBI astrometry of W3(OH) relative to the phase-reference sources J0223+6307 and J0244+6228. The contributions from Earth orientation parameters (EOP), tropospheric delay residuals, ionospheric delay residuals, station position uncertainties, instrumental delay calibration errors, and thermal noise are shown separately. The angular separations between W3(OH) and the phase-reference sources are indicated in parentheses below the source names. For the thermal noise contribution, signal-to-noise ratios of 16 and 35 were assumed for the W3(OH)–J0223+6307 and W3(OH)–J0244+6228 pairs, respectively. {Alt text: Bar chart showing the astrometric position error budget for W3(OH) relative to J0223+6307 and J0244+6228. Error contributions from EOP, troposphere, ionosphere, station position, instrumental effects, and thermal noise are compared.}}
\label{fig:11} 
\end{figure}  

\subsection{Phase calibration}
Since the correlator delay model of KJCC is not sufficiently accurate for high-precision astrometry, the total delay derived from the correlator delay model (i.e., $\tau$, $\dot{\tau}$, $\ddot{\tau}$, and $\dddot{\tau}$), together with the clock parameters, was replaced with a more accurate total delay recalculated using the improved delay model described below. The resulting delay corrections were applied during data reduction using the AIPS task $\texttt{TBIN}$.

Using the phase-referencing technique, the observed delay difference between the target and phase-reference sources can be described as (e.g., \citealp{2014ARA&A..52..339R})
\begin{equation}
\begin{split}
\label{Eq:relative-VLBI}
  \tau_{{\rm1}} - \tau_{{\rm2}} \ &= \  (\tau_{{\rm geo, 1}} - \tau_{{\rm geo, 2}}) + (\tau_{{\rm tropo, 1}} - \tau_{{\rm tropo, 2}}) \\ &+ (\tau_{{\rm iono, 1}} - \tau_{{\rm iono, 2}}) + (\tau_{{\rm ant, 1}} - \tau_{{\rm ant, 2}})
    + (\tau_{{\rm inst, 1}} - \tau_{{\rm inst, 2}})\\ & + (\tau_{{\rm struc, 1}} - \tau_{{\rm struc, 2}})+ (\tau_{{\rm therm, 1}} - \tau_{{\rm therm, 2}})
    \end{split}
\end{equation}
where subscripts 1 and 2 denote the target and reference sources, respectively. Here, $\tau_{{\rm geo}}$ is the geometric delay, $\tau_{{\rm tropo}}$ the tropospheric delay, $\tau_{{\rm iono}}$ the ionospheric delay, $\tau_{{\rm ant}}$ the delay caused by station position errors, $\tau_{{\rm inst}}$ the instrumental delay, $\tau_{{\rm struc}}$ the delay due to source structure, and $\tau_{{\rm therm}}$ the delay caused by thermal noise. For high-precision astrometry, it is essential to accurately calculate the geometric delay, as it contains the relative position information between the target and reference sources (i.e., the astrometric result). \citet{2020PASJ...72...52N} showed that the position error in VLBI phase-referencing can be approximated as
\begin{equation}
\label{Eq:position-error}
\Delta\theta \approx \frac{c\Delta\tau_{{\rm err}}}{D_{{\rm proj}}} \approx \frac{c\Delta\tau_{{\rm err}}}{\lambda}\theta_{{\rm beam}}
\end{equation}
where $c$ is the speed of light, $\Delta\tau_{{\rm err}}$ is the delay calibration error between the target and a phase reference source, and $D_{{\rm proj}}$ is the projected baseline length, $\lambda$ is the observing wavelength, and $\theta$ is the synthesized beam size.
To calculate accurate geometric delays, it is necessary to minimize delay residuals using accurate geophysical models and observational data, as described in the following subsections.
An error budget for the VLBI astrometric position error arising from the major sources of delay calibration uncertainty is presented in Figure~\ref{fig:11}. Note that the source structure effect is not shown in this figure.
\subsubsection{EOP}
The VERA software for accurate geometric delay calculations uses the Earth Orientation Parameters (EOP) from the IERS 2014 C04 series\footnote{\url{ftp://hpiers.obspm.fr/eop-pc/eop/eopc04_14/eopc04.62-now}}. The IERS 14C04 series is provided by the International Earth Rotation and Reference Systems Service (IERS). Differences between the IERS 14C04 series and the IVS or IGS combined solutions show standard deviations of $\mathrm{40\,\mu as}$, $\mathrm{10\,\mu as}$, and $\mathrm{30\,\mu as}$ for nutation, UT1, and polar motion, respectively \citep{2019JGeod..93..621B}. Since errors in the EOP introduce orientation errors in the VLBI baselines, the resulting position error (Equation \ref{Eq:position-error}) can be approximated as
\begin{equation}
\label{Eq:position-error_EOP}
\Delta\theta \approx \left(\frac{c}{D_{\rm proj}}\right)\left(\frac{D_{\rm proj}\,\Phi_{\rm err}\,|\mathbf{s}_{1}-\mathbf{s}_{2}|}{c}\right) \approx 2\ \left(\frac{\Phi_{\rm err}}{50\ \mu{\rm as}}\right)\left(\frac{\theta_{\rm SA}}{2^{\circ}}\right)\ \mathrm{[}\mu\mathrm{as]}
\end{equation}
where $\Phi_{\rm{err}}$ is the EOP error, $\theta_{{\rm SA}}$ is the separation angle between the target and the phase-reference source, and $\mathbf{s}_1$ and $\mathbf{s}_2$ are unit vectors pointing toward the target and reference sources, respectively.

\subsubsection{Tropospheric calibration}
Tropospheric zenith delays were estimated using three methods: (1) analysis of collocated GPS data, (2) integration of the atmospheric refractivity profile derived from the JMA mesoscale analysis data, and (3) analysis of VLBI data obtained in geodetic blocks. To convert the zenith delay to the line-of-sight tropospheric delay (i.e., tropospheric slant delay), we applied the Niell mapping function (\citealp{1996JGR...101.3227N}) to the GPS and JMA results. For the geodetic-block results, we used the AIPS task \texttt{DELZN}, which assumes a mapping function of $1/\sin(\mathrm{El})$, where El is the antenna elevation angle. For an antenna located at a latitude of 30$^{\circ}$, the ratios between the two mapping functions are 99.8$\%$ and 98.3$\%$ at elevations of 30$^{\circ}$ and 10$^{\circ}$, respectively. The average latitude of the KaVA array is 33$^{\circ}$. In these KaVA observations, the data were obtained at relatively high elevation angles (>30$^{\circ}$); therefore, the difference between the two mapping functions is negligible.

GPS receivers (Trimble NetRS) are installed at all VERA stations, and the GPS data are analyzed using the GIPSY-OASIS II software (version 6.4). The time resolution of the VERA GPS data is 5 minutes, and the zenith wet delays derived from the GPS data typically range from 0.05 to 0.40 m (see \citealp{2020PASJ...72...52N} for details). GNSS receivers (Trimble NetR9) are installed at all KVN stations, and the GNSS data are analyzed using the Bernese GNSS Software (version 5.2). The time resolution of the KVN GNSS data is 15 minutes.

JMA mesoscale analysis data\footnote{\url{https://www.jma.go.jp/jma/jma-eng/jma-center/nwp/outline2026-nwp/index.htm}} are atmospheric fields of temperature, pressure, and humidity with a temporal resolution of 3 hours (assimilation interval), produced by assimilating meteorological observations into the JMA mesoscale numerical weather prediction system. These data have a horizontal grid spacing of 5 km $\times$ 5 km and cover Japan and the surrounding regions, with a domain size of 4,080 $\times$ 3,300 km. Using these atmospheric fields, the tropospheric zenith delay was estimated by calculating the atmospheric refractivity and integrating the resulting refractivity profile along the vertical direction. The JMA data can be applied to all VERA and KVN stations, as well as to some stations of the Chinese VLBI Network (CVN) (e.g., the Tianma 65 m telescope). 

In the geodetic blocks, ICRF sources were observed over a wide range of azimuth and elevation angles within each $\sim$30 minute block. Four geodetic blocks were included in each VLBI observation. We followed the procedures described on the NRAO web page\footnote{\url{https://casaguides.nrao.edu/index.php/AIPS-Spectral_Lines_and_Astrometry#Reducing_geodetic-style_data_.28BM272HC_GEO.UVDATA.1.29}} to determine the tropospheric zenith delays and clock parameters from the geodetic-block data.

\citet{2015PASJ...67...65N} reported a tropospheric zenith delay error of $c\tau_{{\rm err}}$ $\sim$2 cm for both the GPS and JMA cases. \citet{2008PASJ...60..951H} compared three methods for tropospheric delay calibration---namely, (1) GPS, (2) geodetic blocks, and (3) image optimization---and showed that all three methods can achieve an accuracy of $\sim$2 cm in the tropospheric zenith delay. Current VLBI astrometry at 22 GHz is limited by residual tropospheric zenith delay and source structure effects (e.g., \citealp{2014ARA&A..52..339R}), provided that the signal-to-noise ratio of the phase-referenced images is sufficiently high that thermal noise is not the dominant error source. The position error (Equation \ref{Eq:position-error}) due to the tropospheric delay residual can be approximated as
\begin{equation}
\label{Eq:position-error_trop}
\Delta\theta \approx (\frac{c}{D_{\rm proj}})(\tau_{{\rm err}}\Delta \sec Z) \approx 41\ (\frac{\mathrm{2300 \ km}}{D_{\rm proj}})(\frac{c\tau_{{\rm err}}}{\mathrm{2 \ cm}})(\frac{\Delta \sec Z}{0.023}) \ \mathrm{[}\mu\mathrm{as]}
\end{equation}
where $\Delta\sec Z$ is the difference in $\sec Z$ (the secant of the zenith angle) between the target and the phase-reference source. The longest baseline length of KaVA is 2300 km, and the average values of $\Delta \sec Z$ between W3(OH) and J0223+6307 and between W3(OH) and J0244+6228 are 0.015 and 0.023, respectively, in these observations.


\begin{figure*}[tbhp] 
 \begin{center} 
     \includegraphics[scale=1.0]{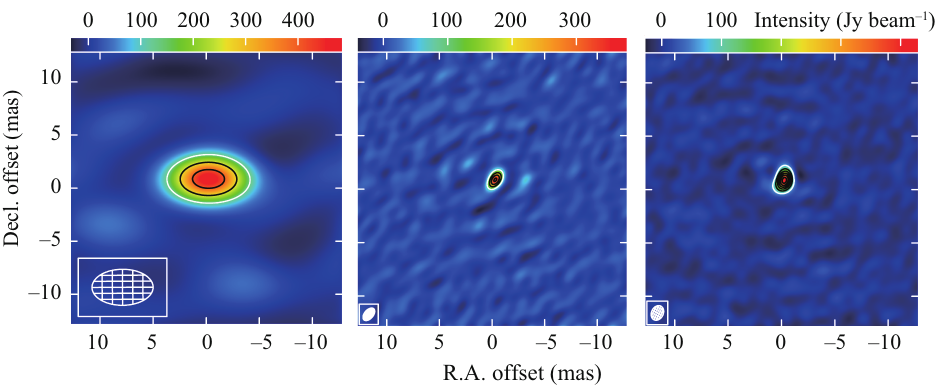} 
\end{center} 
\caption{Phase-referenced images of maser feature ID1 (Table~\ref{table:6}) with respect to J0244+6228. GPS data were used for the tropospheric delay calibration. The images were obtained with KVN (left), VERA (middle), and KaVA (right) on 2017 December 18 (Table~\ref{table:1}). The synthesized beam is shown in the lower-left corner of each panel. Contours are drawn at intervals of 10$\sigma$. The map center is $(\alpha, \delta)_{\rm J2000.0} = (\timeform{02h27m04s.8394}, +\timeform{61D52'24''.608})$. {Alt text: Three phase-referenced images of H$_2$O maser feature ID1 obtained with KVN, VERA, and KaVA. The KaVA image shows higher signal-to-noise ratio than the VERA image.}}
\label{fig:1} 
\end{figure*}  

\begin{figure}[tbhp] 
 \begin{center} 
     \includegraphics[scale=1.0]{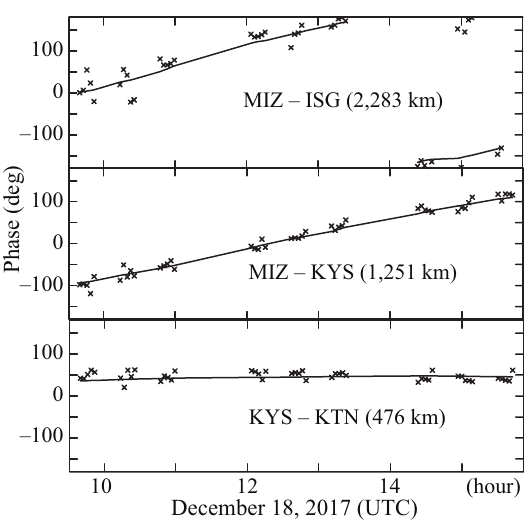} 
\end{center} 
\caption{Calibrated phases from Figure \ref{fig:1} are shown as a function of time (UTC). The top, middle, and bottom panels correspond to different baselines, with the baseline length for each panel indicated in parentheses (MIZ = VERA Mizusawa; ISG = VERA Ishigaki-jima; KYS = KVN Yonsei; KTN = KVN Tamna). Each cross represents the phase obtained by integrating the visibilities over 3 minutes, while the black curves show model phases derived from AIPS Clean Component (CC) files. {Alt text: Three panels showing visibility phase as a function of time for different baselines, including calibrated data points and model curves.}} 
\label{fig:2} 
\end{figure}  

\subsubsection{Ionospheric calibration}

The ionospheric delays were corrected using the CODE (Center for Orbit Determination in Europe) Global Ionosphere Maps (GIM)\footnote{CODE GIM:\url{http://ftp.aiub.unibe.ch/CODE/IONO/}} \citep{2009JGeod..83..263H}. A single-layer model \citep{1999GGAS...59.....S} was assumed to calculate the ionospheric delay in the source direction (i.e., the ionospheric slant delay). The temporal resolution of the CODE GIM has been 1 hour since 2013. The error in the total electron content (TEC$_{\mathrm{err}}$) is 3-10 TEC units \citep{1997RaSc...32.1499H}, and the ionospheric delay residual can be expressed as
\begin{equation}
\label{Eq:ionospheric-delay}
c\tau_{\mathrm{err}} \approx 8.3\left(\frac{22\ \mathrm{GHz}}{\nu}\right)^{2} \left(\frac{\mathrm{TEC_{\mathrm{err}}}}{10\ \mathrm{TECU}}\right)\ {\rm[mm]}
\end{equation}
where $\nu$ is the observing frequency. Using Equation (\ref{Eq:ionospheric-delay}), the position error (Equation \ref{Eq:position-error}) can be approximated as
\begin{equation}
\label{Eq:position-error_iono}
\Delta\theta \approx 13\ (\frac{\mathrm{2300 \ km}}{D_{\rm proj}})\left(\frac{22\ \mathrm{GHz}}{\nu}\right)^{2} \left(\frac{\mathrm{TEC_{\mathrm{err}}}}{10\ \mathrm{TECU}}\right)(\frac{\Delta \sec Z'}{0.017}) \ \mathrm{[}\mu\mathrm{as]}
\end{equation}
where $\Delta\sec Z'$ is the difference in $\sec Z'$ (the ionospheric mapping function) between the target and the phase-reference source. Note that 
\begin{equation}
\sin Z' = \frac{R}{R+H}\sin Z,
\label{eq:zenith}
\end{equation}
where $R \approx 6371\ \mathrm{km}$ is the mean Earth radius and $H = 450\ \mathrm{km}$ is the height of the single-layer ionosphere above the Earth's surface \citep{1999GGAS...59.....S}. The average values of $\Delta \sec Z'$ between W3(OH) and J0223+6307 and between W3(OH) and J0244+6228 are 0.011 and 0.017, respectively, in these observations.


\subsubsection{Antenna position errors}
Antenna positions of VERA and KVN have been regularly monitored using GPS and 22 GHz geodetic observations since 2014, and the relative antenna positions within the KaVA array have been determined with an accuracy of $\sim$3 mm (the project's internal solution, v2005trf14; \citealp{2018JGSP...63...193}). Since 2020, 22 GHz geodetic observations have been conducted twice per year using the EAVN, which includes VERA, KVN, the Takahagi 32 m telescope, and CVN stations. The position error (Equation \ref{Eq:position-error}) caused by antenna position errors can be approximated as
\begin{equation}
\label{Eq:position-error_Ant}
\Delta\theta \approx \left(\frac{c}{D_{\rm proj}}\right)\left(\tau_{\rm err}\theta_{\rm SA}\right) \approx 9\ \left(\frac{2300\ \mathrm{km}}{D_{\rm proj}}\right)\left(\frac{c\tau_{\rm err}}{3\ \mathrm{mm}}\right)\left(\frac{\theta_{\rm SA}}{2^{\circ}}\right)\ \mathrm{[}\mu\mathrm{as]}
\end{equation}
\subsubsection{Source structure and reference position errors}
Unmodeled source structure introduces additional delays ($\tau_{{\rm struc}}$). The effect of source structure is a baseline-based quantity and can be modeled during data reduction using iterative self-calibration techniques (e.g., \citealp{1995ASPC...82...39C}). We applied iterative self-calibration to our data using the AIPS tasks $\texttt{IMAGR}$ and $\texttt{CALIB}$. 

\citet{2014ARA&A..52..339R} discussed the first- and second-order effects of positional errors of the phase-reference source on the target. The first-order effect is that the position offset of the reference source is directly transferred to the target position. In contrast, second-order effects introduce additional phase errors, resulting in degraded image quality and small position errors. The second-order phase shift $\delta\phi$ can be approximated as

\begin{equation}
\label{Eq:source-position}
\delta\phi \approx 1\left(\frac{\theta_{\rm SA}}{1^{\circ}}\right)\left(\frac{1\ \mathrm{mas}}{\theta_{\rm beam}}\right)\left(\frac{\delta_{\theta}}{10\ {\rm mas}}\right) \ {\rm[rad]}
\end{equation}
where $\delta_{\theta}$ is the positional error of the reference source. This effect cannot be directly translated into an astrometric position error in VLBI phase-referencing (Equation \ref{Eq:position-error}). \citet{2014ARA&A..52..339R} suggested that the positional accuracy of the reference source should be better than 10 mas to minimize this effect. However, positional accuracies of our calibrators, J0223+6228 and J0244+6228, are more than two order of magnitude better than this requirement \citep{2025ApJS..276...38P}. 

\subsubsection{Instrumental delays}
Time variations in the parallactic angle were corrected using the AIPS task $\texttt{CLCOR}$. Note that this correction was not applied to VERA, because VERA employs a dual-beam system \citep{2000SPIE.4015..544K}, in which a field rotator compensates for the parallactic angle during observations. The delay difference introduced by the VERA dual-beam system was calibrated using the ``horn-on-dish'' method \citep{2008PASJ...60..935H}, in which common wideband noise is injected into the dual-beam receivers to monitor the path-length difference between the two beams. The horn-on-dish method is designed to calibrate the instrumental delay difference between the two beams of a VERA antenna and therefore does not compensate for instrumental phase or delay differences between VERA and the KVN stations. The dual-beam delay error in the antenna structure and receiver is $c \tau_\mathrm{err} \approx$ 0.1 mm \citep{2008PASJ...60..935H}. Since the residual dual-beam delay originates in the instrumental signal path rather than in differential atmospheric propagation, the corresponding astrometric position error can be approximated by Equation~(\ref{Eq:position-error}) as
\begin{equation}
\label{Eq:position-error_Ins}
\Delta\theta \approx 9\ (\frac{2300 \ \mathrm{km}}{D_{\rm proj}})(\frac{c\tau_{{\rm err}}}{0.1 \ \mathrm{mm}})\ \mathrm{[}\mu\mathrm{as].}
\end{equation}
Note that an independent calibration table is generated for the dual-beam calibration and can be loaded, after standard phase-referencing, using the AIPS task \texttt{TBIN}.


\subsubsection{Thermal error}
The astrometric position error (Equation \ref{Eq:position-error}) due to thermal noise can be expressed as (see \citealp{2014ARA&A..52..339R})
\begin{equation}
\label{Eq:thermal-error}
\Delta\theta \ \approx \ 0.5\left( \frac{\ \theta_{\mathrm{beam}}}{\mathrm{S/N}} \right) \approx 31 \left( \frac{22\ \mathrm{GHz}}{\nu} \right)
\left(
\frac{2300\ \mathrm{km}}{D_{\mathrm{proj}}} \right) \left(
\frac{20}{\mbox{S/N}} \right)
\mbox{[$\mu$as]}
\end{equation}
where S/N is the signal-to-noise ratio of the phase-referenced image. 

\section{Results}

\begin{sidewaystable*}[htbp]
\caption{Phase-referencing and parallax/proper-motion fitting results for W3(OH). } 
\begin{center} 
\label{table:6} 
\begin{tabular}{cccccccccccccc} 
\hline 
\hline 
 ID&\multicolumn{2}{c}{Position offset}&<$V_{\rm{LSR}}$>&\multicolumn{2}{c}{$S_{\rm peak}$}&S/N$_{\rm{med}}$&Parallax          &\multicolumn{2}{c}{Proper motion}&Phase &Tropospheric &\multicolumn{2}{c}{Post-fit residuals}
 \\
 \cline{2-3}\cline{5-6}\cline{9-10}\cline{13-14} 
 &<$\alpha \mathrm{cos} \delta$>&<$\delta$>&&Min&Max&&&$\mu_{\alpha} \cos\delta$&$\mu_{\delta}$&reference&calibration&$\Delta\alpha \mathrm{cos} \delta$&$\Delta\delta$\\ 
   &(mas)&(mas)&(km s$^{-1}$)&\multicolumn{2}{c}{(Jy beam$^{-1}$)}&  &(mas)&(mas yr$^{-1}$)&(mas yr$^{-1}$)&    &  &(mas) &(mas)\\
  \hline

1&0&1& $-$47.7&346&2321&35&0.490$\pm$0.028      &0.78$\pm$0.06        &1.38$\pm$0.24                                &J0244+6228&GPS           &0.036&0.181 \\ 

1&0&1&   //&166&2174&  29&0.476$\pm$0.044    &0.83$\pm$0.09        &1.26$\pm$0.26                   &//&JMA   &0.055&0.186\\ 

2&13&29&$-$50.2&20&600& 18&0.461$\pm$0.043    &$-$0.29$\pm$0.08        &1.83$\pm$0.15                                  &//&GPS      &0.055&0.115      \\ 

2&13&29&   //&12&525& 16&0.440$\pm$0.056    &$-$0.18$\pm$0.11        &1.70$\pm$0.17                               &//&JMA      &0.070&0.120      \\

3&$-$1935&24& $-$81.9&4&40& 35&0.457$\pm$0.036  &1.82$\pm$0.07        &$-$3.21$\pm$0.26             &//&GPS      &0.045&0.198     \\

3&$-$1935&24&  //&6&37&   29&0.438$\pm$0.031     &1.84$\pm$0.06        &$-$3.31$\pm$0.29                              & // &JMA      &0.037&0.210     \\

4&$-$2615&$-$206& $-$51.8&4&26&  24&0.449$\pm$0.154    &$-$3.60$\pm$0.37        &$-$2.55$\pm$0.30                                 &//&GPS      &0.252&0.211     \\

4&-2615&$-$206&  $-$51.9 &6&13&  20&0.607$\pm$0.142     &$-$3.69$\pm$0.47        &$-$2.68$\pm$0.22                             &//&JMA      &0.328&0.150     \\

5&$-$1097&$-$44& $-$58.8&3&9&  15&0.478$\pm$0.055   &$-$2.75$\pm$0.14        &$-$1.43$\pm$0.12                   &//&GPS      &0.085&0.081   \\

5&$-$1097&$-$44&  //&3&9&    15&0.476$\pm$0.047  &$-$2.74$\pm$0.12        &$-$1.62$\pm$0.11               &//&JMA      &0.059&0.078   \\

6&$-$1116&$-$42&$-$60.3&2&29& 36&0.530$\pm$0.017       &$-$3.91$\pm$0.04        &$-$1.43$\pm$0.07               &//&GPS      &0.023&0.052   \\

6&$-$1117&$-$42&  //&2&31&   28&0.514$\pm$0.019    &$-$3.94$\pm$0.05        &$-$1.51$\pm$0.06             &//&JMA      &0.025&0.048   \\


1&0&2& $-$47.7&83&1834& 16&0.494$\pm$0.052   &0.59$\pm$0.11        &1.70$\pm$0.24     &J0223+6307&GPS      &0.051&0.180     \\ 

1&0&2&    //&71&1160& 16    &0.525$\pm$0.050    &0.56$\pm$0.10        &1.67$\pm$0.28        &//&JMA   &0.053&0.204\\ 


2&13&30&  $-$50.2 &5&436&7&0.516$\pm$0.058   &$-$0.36$\pm$0.10        &1.90$\pm$0.21               &//&GPS      &0.055&0.185\\ 

2&13&30& $-$50.1&8&271&   11&0.463$\pm$0.072    &$-$0.45$\pm$0.15        &2.00$\pm$0.18           &//&JMA      &0.114&0.153\\


3&$-$1935&25&  $-$81.9 &1&26&    15&0.537$\pm$0.098   &1.57$\pm$0.21        &$-$2.93$\pm$0.32        &//&GPS      &0.125&0.226     \\

3&$-$1935&25&   //&1&22&   14 &0.700$\pm$0.120     &1.40$\pm$0.30        &$-$2.97$\pm$0.23        &//&JMA      &0.199&0.162     \\



5&$-$1097&$-$43& $-$60.3&1&4&  10&0.543$\pm$0.082    &$-$2.99$\pm$0.18        &$-$1.02$\pm$0.26     &//&JMA      &0.100&0.190   \\


  \hline

& &&&&&&Combined fit \\
& &&&   &              && {\bf 0.497$\pm$0.009}\footnotemark[$\ast$]      &                 \\ 
\hline

\multicolumn{4}{@{}l@{}}{\hbox to 0pt{\parbox{215mm}{\normalsize
\par\noindent
\\
Column 1: Maser feature ID; Columns 2--3: Mean position offsets in right ascension and declination, respectively, averaged over the six observing epochs and measured relative to $(\alpha, \delta)_{\rm J2000.0} = (\timeform{02h27m04s.8394}, +\timeform{61D52'24''.608})$;  Column 4: Mean LSR velocity of the maser feature; Columns 5--6: Minimum and maximum peak flux densities of the maser feature, respectively; Column 7: Median signal-to-noise ratio of the phase referenced images; Column 8: Parallax results; Columns 9--10; Proper motion components in the east and north directions, respectively; Column 11: Phase-reference source; Column 12: Tropospheric calibration methods. JMA refers to the Japan Meteorological Agency mesoscale analysis data (see text for details); Columns 13--14: Root-mean-square (RMS) values of the post-fit residuals in right ascension and declination, respectively.\\
\footnotemark[$\ast$] The final parallax is adopted as 0.497 $\pm$ 0.024 mas because the formal uncertainty of the parallax is underestimated (see the text for details). 

}\hss}}
\end{tabular} 
\end{center} 
\end{sidewaystable*}

\begin{figure*}[tbhp] 
 \begin{center} 
     \includegraphics[scale=1.0]{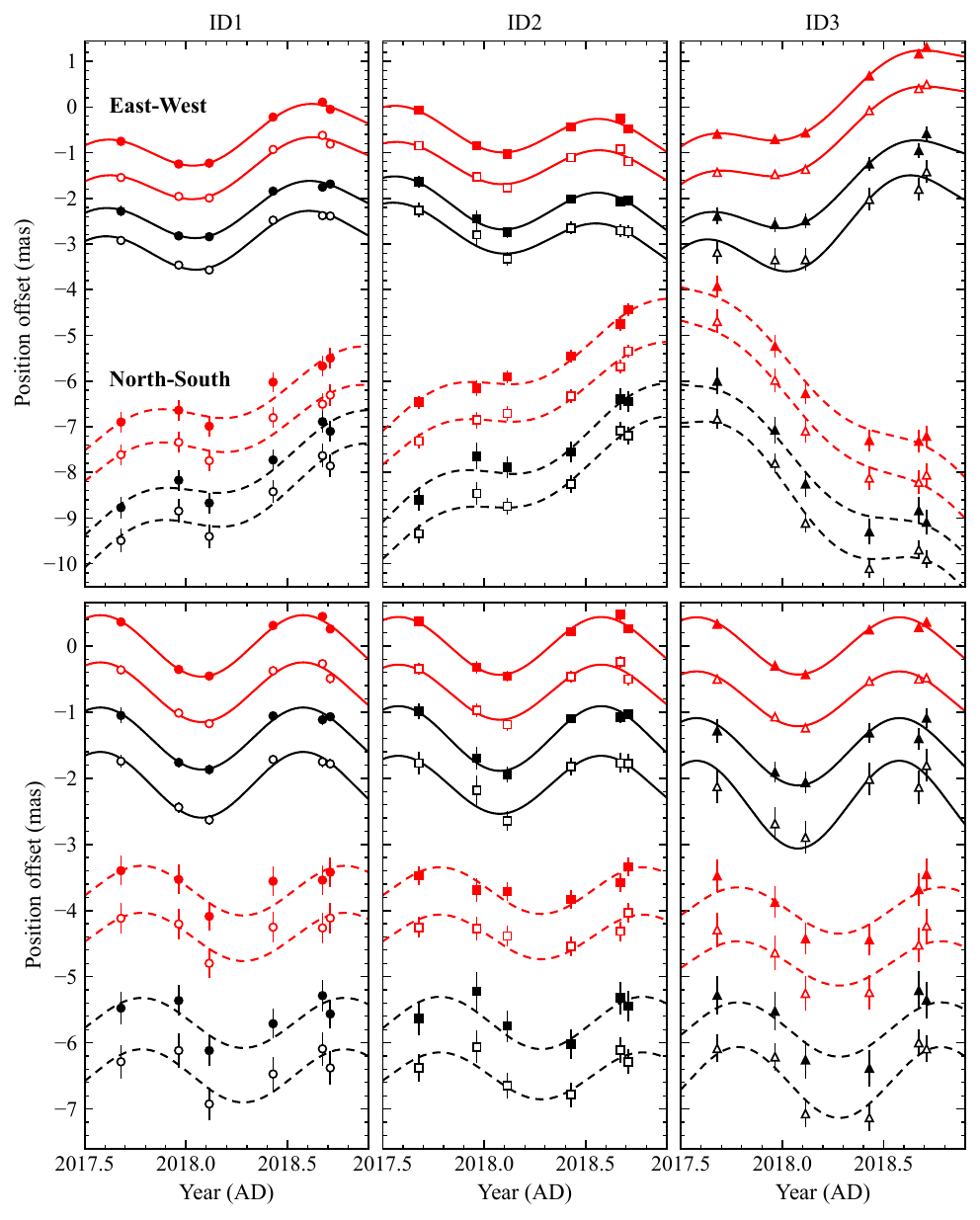} 
\end{center} 
\caption{{\bf(Top row)} Trigonometric parallax and proper motion fitting results for maser features ID1 (left), ID2 (middle), and ID3 (right) (see Table \ref{table:6}). Solid and dashed curves represent the best-fit models in the east ($\alpha\cos\delta$) and north ($\delta$) directions, respectively. Red and black symbols indicate the results obtained using the phase-reference sources J0244+6228 and J0223+6307,
respectively. Filled and open symbols denote the tropospheric calibrations based on GPS and JMA data, respectively. For clarity, the datasets and their corresponding model curves are vertically offset for display purposes only. {\bf(Bottom row)} Same as the top row, but with the fitted proper motions removed, showing only the parallax component. {Alt text: Panels showing position offsets as a function of time for maser features ID1, ID2, and ID3. The top row shows the full astrometric fits, while the bottom row shows the parallax component after removing the fitted proper motions.}}
\label{fig:4} 
\end{figure*}  

\subsection{Phase-referencing results}

Figure~\ref{fig:1} shows phase-referenced images of the H$_2$O maser feature ID1 (Table~\ref{table:6}) obtained with KVN, VERA, and KaVA (from left to right). The image signal-to-noise ratios (S/N) are 36, 42, and 65 for the KVN, VERA, and KaVA images, respectively. The S/N of the KaVA image is 55$\%$ higher than that of the VERA image, in good agreement with the theoretical expectation of $\sim$60$\%$. The theoretical expectation is estimated as $\sigma_{\rm im} = \left( \sum \sigma_{\rm bl}^{-2} \right)^{-1/2}$, where $\sigma_{\rm bl}$ is the thermal noise level for an individual interferometric baseline. Each baseline noise level was estimated from the typical SEFD (System Equivalent Flux Density) values of the individual telescopes. Figure \ref{fig:2} shows the calibrated visibility phases corresponding to the KaVA image in Figure \ref{fig:1} for different baseline lengths. The phase calibration of the KaVA data is successful over all baselines. 

However, we found systematic delay offsets between the VERA and KVN arrays in the FRING solutions for the background continuum sources J0223+6307 and J0244+6228 after applying the parallactic-angle correction, the delay recalculation table described above, and the clock parameter calibration. We confirmed that these delay offsets disappear when VERA operates in single-beam mode rather than dual-beam mode (see Figure \ref{fig:3} in Appendix). Similar results to those shown in Figure~\ref{fig:3} were also found at other observing epochs. The delay offset is approximately constant over time and depends on the target--reference separation angle in the VERA dual-beam system. Phase referencing was successful when a continuum source was used as the phase reference. In contrast, it failed when a spectral-line source (i.e., W3(OH)) was used as the phase reference. Therefore, we adopted the continuum sources J0223+6307 and J0244+6228 as phase references instead of the spectral-line source W3(OH). This systematic delay offset cannot be corrected when a spectral-line source is used as the phase reference, because such data can only be used to remove rate residuals, not delay residuals. These systematic delay offsets do not affect astrometric observations conducted solely with VERA, because all stations share the same dual-beam system. However, they can become significant when VERA is combined with non-VERA stations, such as KVN stations in KaVA observations. When VERA is operated in single-beam mode, both continuum and spectral-line sources can therefore be used as phase references in future KaVA astrometric observations.

An exception occurred during the first observing epoch. At that epoch, J0223+6307 was relatively faint ($\sim$20\,$\pm$\,3 mJy), and several stations exhibited high system noise temperatures ($T_{\rm sys}^{*} \sim 200$--1100 K). As a result, phase referencing using J0223+6307 was unsuccessful. Therefore, for the analysis associated with J0223+6307 in the first epoch, we used the bright maser feature ID1 ($\sim$500\,$\pm$\,50 Jy) as the phase-reference source and analyzed only the VERA data. For all other epochs, J0223+6307 was used as the phase-reference source to evaluate the astrometric performance of KaVA. In addition, J0244+6228 was adopted as the phase-reference source for all epochs.

\begin{figure*}[tbhp] 
 \begin{center} 
     \includegraphics[scale=1.0]{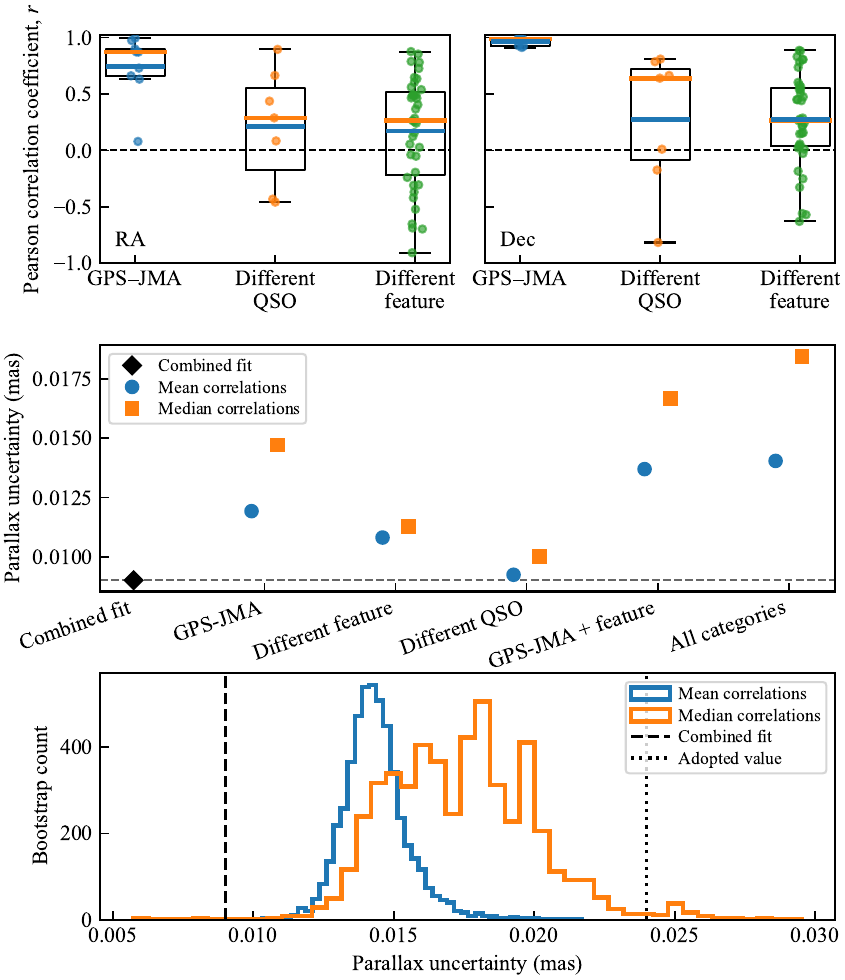} 
\end{center} 
\caption{{\bf(Top)} Empirical distributions of the post-fit residual correlation coefficients. The box plots summarize the pairwise Pearson correlation coefficients of the post-fit residuals for the calibration-, feature-, and quasar-related residuals in right ascension (left) and declination (right). Blue and orange symbols indicate the corresponding mean and median correlation coefficients, respectively. {\bf(Middle)} Comparison of the formal parallax uncertainties obtained with different covariance models.
Blue circles and orange squares represent the formal uncertainties obtained using covariance matrices constructed from the mean and median empirical correlation coefficients, respectively. The black horizontal dashed line indicates the formal parallax uncertainty from the combined fit listed in Table~\ref{table:6}. {\bf(Bottom)} Bootstrap distributions of the formal parallax uncertainties obtained by propagating the uncertainties in the post-fit residual correlation coefficients. The blue and orange histograms show the distributions of the formal parallax uncertainties obtained using the mean and median post-fit residual correlation coefficients for each bootstrap realization, respectively. The black dashed and dotted vertical lines indicate the combined-fit formal parallax uncertainty (Table~\ref{table:6}) and the adopted parallax uncertainty, respectively. {Alt text: Composite figure showing distributions of post-fit residual correlation coefficients for calibration, maser-feature, and quasar categories in right ascension and declination, comparison of formal parallax uncertainties under different covariance assumptions, and bootstrap distributions of the resulting parallax uncertainties.}}
\label{fig:13} 
\end{figure*}

\subsection{Parallax and proper motion results}
\label{Section:4-2}


The parallax and proper-motion were determined through weighted least-squares fitting, using the reference positions ($\alpha_{0}$, $\delta_{0}$), proper motion components ($\mu_{\alpha} \cos\delta$, $\mu_{\delta}$), and parallax $\pi$ as free parameters. The geodetic-block analysis failed due to residual instrumental delays associated with the VERA dual-beam mode, since multi-band delays are modeled together with clock and atmospheric terms in the AIPS task \texttt{DELZN}. Following previous studies (e.g., \citealp{2019ApJ...885..131R}), error floors were added in quadrature to the formal measurement uncertainties (i.e., statistical errors) such that the reduced chi-square values of the fits became close to unity. Thus, the final error floors in right ascension and declination for each fit can be estimated from the post-fit residuals listed in Table \ref{table:6}, multiplied by $\sqrt{N/(N-2)}$ = $\sqrt{3/2}$, where $N$ is the number of data points (i.e., $N$ = 6). The origin of these error floors (i.e., systematic errors) is likely attributable to tropospheric delay residuals and/or source structure effects, which are known to be significant in VLBI astrometry at observing frequencies above 10 GHz (e.g., \citealp{2020PASJ...72...52N}; \citealp{2014ARA&A..52..339R}). 

In Table \ref{table:6}, a total of 19 parallax measurements were obtained using two phase-reference sources, two tropospheric calibration methods, and six maser features (i.e., $2 \times 2 \times 6 - 5 = 19$). The signal-to-noise ratios of the phase-referenced images were lower when J0223+6307 was used as the phase-reference source than when J0244+6228 was used. As a result, five of the 24 possible parallax measurements could not be successfully determined. This is likely because J0223+6307 (21\,$\pm$\,4 to 130\,$\pm$\,10 mJy) was significantly fainter than J0244+6228 (440\,$\pm$\,40 to 1000\,$\pm$\,100 mJy) during the observations. Figure \ref{fig:4} shows examples of the parallax and proper-motion fitting results for maser features ID1, ID2, and ID3. The corresponding results for features ID4, ID5, and ID6 are shown in Figure \ref{fig:4b} in the Appendix. 

The combined fit using the 19 data sets yields a parallax of $0.497 \pm 0.009$ mas. However, this uncertainty is likely underestimated because of correlations among the data sets. If all 19 data sets were fully correlated, the uncertainty would increase to $0.009 \times \sqrt{19} = 0.039$ mas, which can be regarded as a conservative upper bound. In previous studies, parallax measurements of different maser features have been considered to be correlated, because residual tropospheric delays affect all maser features in a similar way. In contrast, measurements using different phase-reference sources have been treated as independent, because they correspond to different sky directions. The results obtained using the GPS and JMA tropospheric calibration data may also be partially correlated. Therefore, as a conservative estimate, we multiplied the formal uncertainty by a factor of $\sqrt{2 \times 6 - 5}$, yielding a final parallax of $0.497 \pm 0.024$ mas. This adopted uncertainty corresponds to the assumption that the results obtained using different atmospheric calibration methods (JMA and GPS) and different maser features are fully correlated, whereas those obtained using different background QSOs are independent. The validity of this assumption is examined in the next section. This parallax corresponds to a distance of $2.01^{+0.10}_{-0.09}$ kpc. The obtained parallax is consistent within the uncertainties with the previous VLBA result of $0.489 \pm 0.017$ mas \citep{2006ApJ...645..337H}.


\section{Discussion}
\subsection{Validation of the Parallax Uncertainty Estimate}

To validate the parallax uncertainty estimate, we perform a generalized least-squares (GLS) analysis using the post-fit residuals from the parallax and proper-motion fits, as summarized in Figure~\ref{fig:13}. The GLS analysis is implemented in \texttt{Python} using the linear algebra routines provided by \texttt{NumPy} \citep{harris2020array}.

First, we calculate Pearson correlation coefficients separately for right ascension and declination using pairs of post-fit residual time series grouped into three categories: different background QSOs, different maser features, and different atmospheric calibration methods. Interestingly, the correlation coefficients for the two atmospheric calibration methods (GPS and JMA) are consistently high in both right ascension and declination, with mean values exceeding 0.75. This strongly suggests that the astrometric results obtained with the two atmospheric calibration methods are not independent. In contrast, the correlation coefficients for different background QSOs and different maser features show much larger scatter.

Second, we construct the covariance matrix used in the GLS analysis from the estimated Pearson correlation coefficients. For each pair of astrometric position measurements, the covariance matrix elements are defined as

\begin{equation}
\label{Eq:GLS-covariance-matrix}
C_{ij} =
\begin{cases}
\sigma_i^2, & i=j,\\
\rho_{ij}\sigma_i\sigma_j, & i\neq j,
\end{cases}
\end{equation}
where $\sigma_i$ is the uncertainty of the $i$th astrometric position measurement and $\rho_{ij}$ is the Pearson correlation coefficient assigned according to the corresponding category (different atmospheric calibration methods, different maser features, or different background QSOs). When a given pair of measurements satisfies more than one category, the corresponding Pearson correlation coefficient is taken as the product of the correlation coefficients for those categories. If $\rho_{ij}=0$ for all $i\neq j$, the covariance matrix reduces to that used in conventional weighted least-squares (WLS) analysis. The parallax uncertainty is then estimated from the covariance matrix of the GLS solution,
\begin{equation}
\label{Eq:GLS-pi-uncertainty}
\sigma_{\pi}
=
\sqrt{
\left[
\left(\mathbf{A}^{\mathrm T}\mathbf{C}^{-1}\mathbf{A}\right)^{-1}
\right]_{\pi\pi}
},
\end{equation}
where $\mathbf{A}$ is the design matrix containing the parallax factor, position offsets, and linear proper-motion terms, $\mathbf{C}$ is the covariance matrix defined in Equation \ref{Eq:GLS-covariance-matrix}, and $[\ ]_{\pi\pi}$ denotes the diagonal element corresponding to the parallax parameter. For the present dataset, the covariance matrix $\mathbf{C}$ has dimensions of $228\times228$, corresponding to 19 astrometric data sets (Table~\ref{table:6}) observed at six epochs in both right ascension and declination. The design matrix $\mathbf{A}$ has dimensions of 228$\times$77, where the 77 parameters comprise one common parallax parameter and four astrometric parameters (position offsets and proper motions in right ascension and declination) for each of the 19 astrometric data sets. Using covariance matrices constructed from the estimated Pearson correlation coefficients, Equation~\ref{Eq:GLS-pi-uncertainty} yields parallax uncertainties ranging from 0.009 to 0.018 mas (middle panel of Figure~\ref{fig:13}), depending on whether the mean or median correlation coefficients are adopted.

\begin{figure*}[tbhp] 
 \begin{center} 
     \includegraphics[scale=1.0]{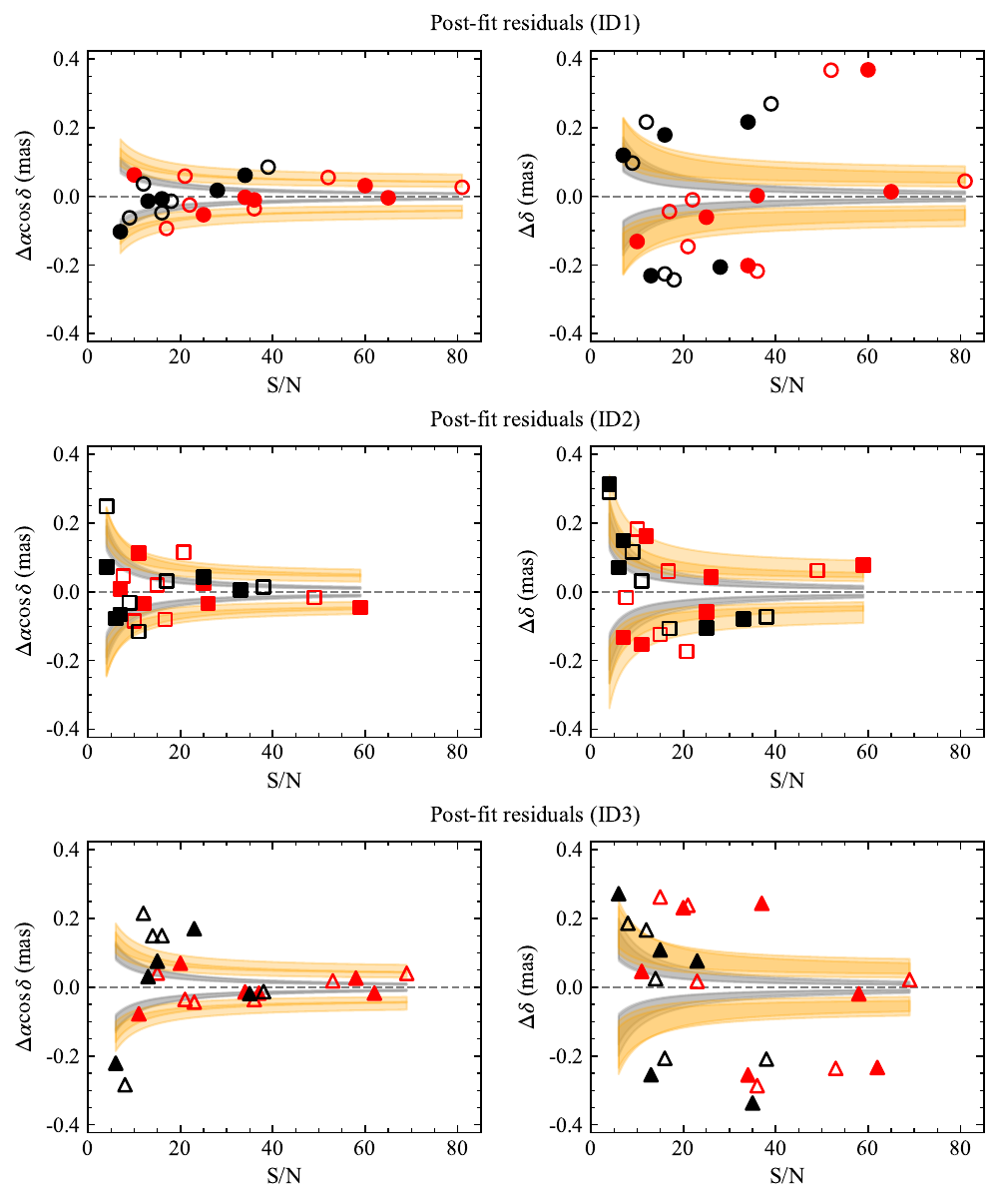} 
\end{center} 
\caption{Post-fit residuals of the parallax and proper-motion fits for maser features ID1 (top row; see Table \ref{table:6}), ID2 (middle row), and ID3 (bottom row) are shown as a function of the signal-to-noise ratio (S/N) of the phase-referenced images. The left and right panels display the residuals in right ascension and declination, respectively. Black and red symbols indicate the results obtained using the phase-reference sources J0223+6307 and J0244+6228, respectively. Filled and open symbols denote the tropospheric calibrations based on GPS and JMA data, respectively. The gray shaded region represents the theoretical position error due to thermal noise (Equation \ref{Eq:thermal-error}), while the orange shaded region represents the root-sum-square of the position errors due to thermal noise and tropospheric delay residuals (Equation \ref{Eq:position-error_trop}). {Alt text: Panels showing the post-fit residuals in right ascension (left column) and declination (right column) for maser features ID1 (top row), ID2 (middle row), and ID3 (bottom row) as a function of the S/N of the phase-referenced images.}} 
\label{fig:6} 
\end{figure*}  

Third, to evaluate the effect of the uncertainty in the Pearson correlation coefficients used to construct the covariance matrix, we perform a bootstrap analysis. In each bootstrap realization, the Pearson correlation coefficients are resampled and the parallax uncertainty is re-estimated using the same GLS procedure. For this bootstrap analysis, the covariance matrix is constructed by considering all three correlation categories (different atmospheric calibration methods, different maser features, and different background QSOs). As shown in the bottom panel of Figure \ref{fig:13}, most bootstrap realizations yield parallax uncertainties between 0.011 and 0.024 mas. This demonstrates that the adopted parallax uncertainty of 0.024 mas used in this paper is conservative. Accordingly, we adopt the parallax of 0.497\,$\pm$\,0.024 mas derived in Section \ref{Section:4-2} as the final parallax of W3(OH).

Furthermore, taking the root-sum-square of the individual contributions to the astrometric error budget shown in Figure \ref{fig:11} yields a single-epoch astrometric error of approximately 50~$\mu$as for the phase-referencing observations with either QSO. With six observing epochs, the expected parallax uncertainty is approximately 20~$\mu$as, which is in good agreement with the adopted final parallax uncertainty of 24~$\mu$as. It should be noted that, although \citet{2007PASJ...59..889H} reported a parallax uncertainty of 8~$\mu$as for S269 using VERA alone, the astrometric accuracy achieved with VERA is known to depend on the target source and observing conditions (e.g., \citealp{2020PASJ...72...50V}).

\subsection{The main cause of post-fit residuals in parallax and proper-motion measurements}
Figure \ref{fig:6} shows the post-fit residuals from the parallax and proper-motion fits for maser features ID1, ID2, and ID3 as a function of the S/N of the phase-referenced images. Similar plots for the other maser features are presented in Figure \ref{fig:6b} in the Appendix. The gray shaded regions indicate the theoretical position errors due solely to thermal noise (Equation \ref{Eq:thermal-error}), while the orange shaded regions represent the root-sum-square of the position errors due to thermal noise and tropospheric delay residuals (Equation \ref{Eq:position-error_trop}). In evaluating Equation~\ref{Eq:position-error_trop}, we adopted a tropospheric delay calibration error (c$\tau_{{\rm err}}$) of 2 cm. For Equations \ref{Eq:position-error_trop} and \ref{Eq:thermal-error}, the beam sizes ($\theta_{\mathrm{beam}}$) in right ascension and declination at each observing epoch were derived from two-dimensional Gaussian fits to the phase-referenced images. 

The orange shaded regions in Figures \ref{fig:6} and \ref{fig:6b} generally provide upper bounds on the post-fit residuals in right ascension, except for maser features ID3 and ID4. For ID3, deviations from the orange shaded region are seen in the measurements obtained using J0223+6307. Since J0223+6307 (21\,$\pm$\,4 to 130\,$\pm$\,10 mJy) was significantly fainter than J0244+6228 (440\,$\pm$\,40 to 1000\,$\pm$\,100 mJy), lower S/N values were obtained in the phase-referenced images when J0223+6307 was used. Therefore, these deviations may be attributable to degraded phase-referencing performance. For ID4, post-fit residuals of approximately $\sim$0.3 and $\sim$0.2 mas are observed in right ascension and declination, respectively, largely independent of the S/N values. This behavior suggests that the astrometric measurements of ID4 may be affected by source structure effects. 

For the post-fit residuals in declination, whether the orange shaded regions provide upper-bounds on the residuals depends on the individual maser feature. No clear systematic difference between the JMA- and GPS-based measurements can be identified from Figures \ref{fig:6} and \ref{fig:6b}.

In summary, to avoid large post-fit residuals (e.g., $>$ 0.1 mas), the phase-referenced images must have sufficiently high S/N values (e.g., S/N $>$ 10). In addition, compact maser sources should be selected for accurate astrometry. Once the S/N exceeds a certain level, the astrometric post-fit residuals are governed by other factors, such as tropospheric delay residuals.

\begin{figure}[tbhp] 
 \begin{center} 
     \includegraphics[scale=1.0]{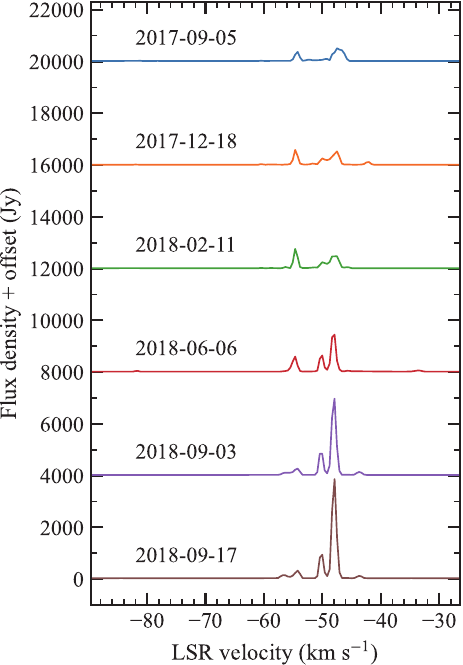} 
\end{center} 
\caption{Scalar-averaged cross-power spectra of W3(OH), averaged over all baselines, obtained during the observations. For clarity, each spectrum is plotted with a vertical offset. The observation date for each spectrum is indicated in the format yyyy--mm--dd. {Alt text: Multiple spectra of 22 GHz H$_{2}$O maser emission obtained at different observation dates.}}
\label{fig:7} 
\end{figure}

\begin{figure}[tbhp] 
 \begin{center} 
     \includegraphics[scale=1.0]{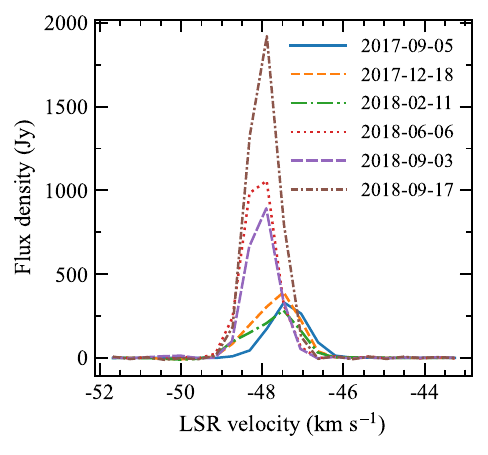} 
\end{center} 
\caption{Results of the AIPS task \texttt{ISPEC} for maser feature ID1 (see Table \ref{table:6}) are shown. The \texttt{ISPEC} task extracts a one-dimensional spectrum by integrating the flux density over a specified region in the image cube as a function of velocity. Different line styles indicate different observation dates. {Alt text: Time-series spectra of maser feature ID1.}}
\label{fig:8} 
\end{figure}  

\begin{figure}[tbhp] 
 \begin{center} 
     \includegraphics[scale=1.0]{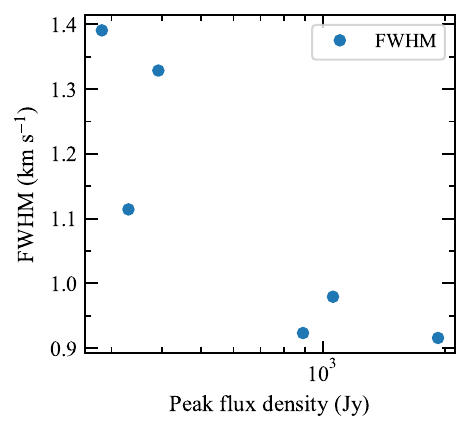} 
\end{center} 
\caption{Line widths are shown as a function of peak flux density for the spectra presented in Figure \ref{fig:8}. The peak flux density is plotted on a logarithmic scale. Blue circles represent the full width at half maximum (FWHM). The Pearson and Spearman rank correlation coefficients are $r = -0.86$ ($p = 0.03$) and $\rho = -0.89$ ($p = 0.02$), respectively. {Alt text: Scatter plot of FWHM line width versus peak maser flux density, showing a strong inverse relationship in which brighter maser emission is associated with narrower line widths.}}
\label{fig:10} 
\end{figure} 

\begin{figure*}[tbhp] 
 \begin{center} 
     \includegraphics[scale=0.95]{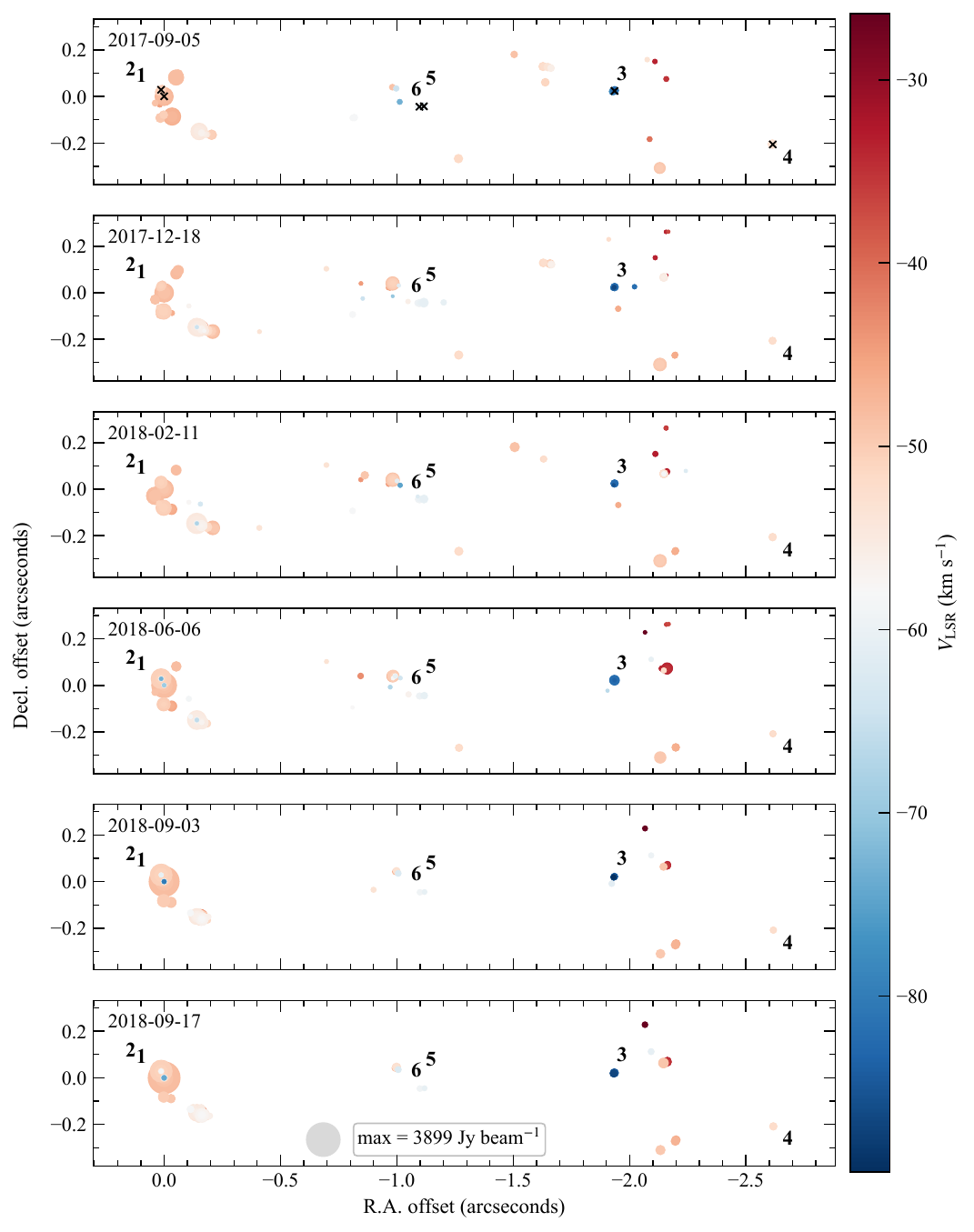} 
\end{center} 
\caption{Spatial distributions of H$_2$O maser spots in W3(OH) at six observing epochs. Colors indicate the LSR velocity, and observation dates are given in each panel. Only maser spots with S/N $>$ 10 are shown. Circle sizes are proportional to the square root of the peak intensity. To maximize spot detection, the maps were produced using maser feature ID1 (Table~\ref{table:6}) as the phase-reference feature. Numbers indicate maser feature IDs, and crosses in the top panel mark their positions. {Alt text: Six panels showing the spatial distribution of H$_2$O maser spots in W3(OH) at different observing epochs. Colors indicate LSR velocity, circle sizes represent peak intensity, and labeled maser features are identified by their feature IDs. Maser features ID1 and ID2 show pronounced brightening during the final three observing epochs.}}
\label{fig:9} 
\end{figure*}  

\begin{figure*}[hptb] 
 \begin{center} 
     \includegraphics[scale=1.0]{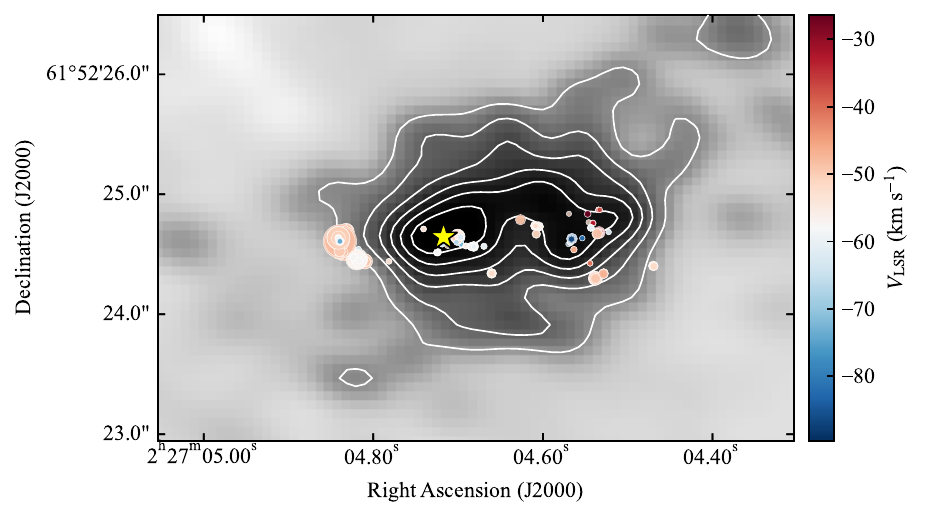} 
\end{center} 
\caption{The NOEMA (IRAM NOrthern Extended Millimeter Array) 1.37 mm (219 GHz) continuum emission toward the W3(OH) region, observed in the ABD configuration, is shown in grayscale and white contours \citep{2018A&A...618A..46A}. The contour levels start at $5\sigma$ and increase in steps of $5\sigma$, where $1\sigma = 3.1$ mJy beam$^{-1}$. The H$_2$O maser spots shown in Figure~\ref{fig:9} are superposed on the continuum map. The yellow star marks the peak position of the VLA 8.4 GHz continuum emission adopted from \citet{2006ApJ...645..337H} and \citet{1999ApJ...513..775W}. {Alt text: Grayscale and contour map of the 1.37 mm continuum emission toward W3(OH), overlaid with H$_2$O maser spots color-coded by their LSR velocity and the VLA 8.4 GHz continuum peak position marked by a yellow star.}}
\label{fig:12} 
\end{figure*}  

\subsection{Flux variation of W3(OH)}
 The peak flux density measured from the cross-power spectrum of W3(OH) at each observing epoch ranged from 510\,$\pm$\,50 to 3900\,$\pm$\,400 Jy (Figure~\ref{fig:7}). This variation became particularly pronounced during the final three observing epochs and was primarily attributable to maser features ID1 and ID2. Figure~\ref{fig:8} shows the one-dimensional spectra of maser feature ID1, obtained by integrating the flux density over a specified narrow region in the phase-referenced image cube. Note that the flux densities shown in Figure~\ref{fig:8} are likely underestimated because of imperfections in the phase-referencing process. Figure~\ref{fig:10} shows strong negative Pearson and Spearman rank correlations between the linewidths and the logarithm of the peak flux density for the spectra presented in Figure~\ref{fig:8}. The Pearson and Spearman rank correlation coefficients are $r=-0.86$ ($p=0.03$) and $\rho=-0.89$ ($p=0.02$), respectively. Figure~\ref{fig:9} shows the spatial distribution of maser spots in the W3(OH) region during the observations. The spatial extent of each map is in good agreement with previous results \citep{2006ApJ...645..337H}. The individual maps indicate that the observed flux variation is primarily associated with maser features ID1 and ID2, which are separated by only $\sim$32 mas ($\sim$64 AU at a distance of 2.01 kpc). Figure \ref{fig:12} shows the H$_2$O maser spots presented in Figure \ref{fig:9} overlaid on the NOEMA (IRAM NOrthern Extended Millimeter Array) 1.37 mm (219 GHz) continuum map \citep{2018A&A...618A..46A}. The spatial relationship between the maser distribution and the dust continuum emission is consistent with previous studies (e.g., \citealp{2006ApJ...645..337H}).

During unsaturated maser amplification, the peak intensity $I(\nu_{0})$ of a Doppler-broadened line increases as $\exp(\tau_{0})$, while the radiation bandwidth $\Delta \nu$ varies approximately as $\tau_{0}^{-1/2}$ for $\tau_{0}\gg1$, where $\tau_{0}$ is the optical depth at the line center (defined as a positive quantity for amplification; \citealt{1992ASSL..170.....E}). Thus, $I(\nu_{0})$ and $\Delta \nu$ are related as 

\begin{equation}
\label{Eq:line-narrowing}
\Delta \nu \propto [\ln I(\nu_{0})]^{-1/2}.
\end{equation}
This line-narrowing effect is clearly seen in maser feature ID1 (Figure~\ref{fig:10}), whereas it is not evident in the other maser features (Figures~\ref{fig:8b} and \ref{fig:8c} in the Appendix). Importantly, both ID1 and ID2 exhibit significant brightening over the same period within a compact region of only $\sim$64 AU. The simultaneous brightening of these nearby maser features suggests a common local enhancement of the maser amplification conditions. The linewidth--intensity anti-correlation observed in ID1 is consistent with an increase in the effective optical depth under unsaturated maser amplification. In contrast, the absence of a clear linewidth variation in maser feature ID2 suggests that maser spots may respond differently to the same environmental change, depending on their saturation state and internal velocity structure.

\section{Summary}
We present the first trigonometric parallax measurements with KaVA, targeting 22-GHz H$_2$O masers associated with the star-forming region W3(OH). The signal-to-noise ratio (S/N) of the phase-referenced images obtained with KaVA is 55$\%$ higher than that obtained with VERA, in good agreement with the theoretical expectation of $\sim$60$\%$ (Figure~\ref{fig:1}). Parallax measurements were successfully obtained using two tropospheric calibration methods (GPS and JMA) and two phase-reference sources, J0223+6307 and J0244+6228 (Table~\ref{table:6}). This study also presents the first successful parallax measurement based on JMA tropospheric calibration. A combined fit to 19 data sets yields a parallax of 0.497$\pm$0.024 mas, corresponding to a distance of 2.01$^{+0.10}_{-0.09}$ kpc. The derived parallax is consistent within the uncertainties with the previous VLBA measurement of 0.489$\pm$0.017 mas \citep{2006ApJ...645..337H}. The post-fit residuals of the parallax and proper-motion measurements are generally consistent with the root-sum-square of the position errors due to thermal noise and a tropospheric delay calibration error of 2 cm (Figures~\ref{fig:6} and \ref{fig:6b}), provided that the maser source is compact and the phase-referenced images maintain sufficiently high S/N ratios (e.g., S/N $>$ 10) throughout the observations. Future KaVA astrometry is expected to increase the number of reliable trigonometric parallax measurements of 22-GHz H$_2$O masers by enabling observations of sources that are difficult to observe with VERA alone because of their low flux densities and limited $uv$ coverage (e.g., sources at declinations of $\sim 0^{\circ}$). This capability will be particularly important for more distant star-forming regions, whose maser emission is generally fainter and therefore more challenging for VERA-only observations.

The parallax obtained with KaVA is less precise than the previous VLBA measurement. However, the astrometric calibration method presented in this paper is sophisticated and can be directly applied to future EAVN observations. The EAVN provides longer baselines than KaVA, which are expected to improve the achievable astrometric precision. The EAVN is also expanding its collaboration to include Southeast Asian stations \citep{2024evn..conf..177S}. This expansion will enhance the capability to observe low-declination Galactic sources that are difficult to access with the VLBA, thereby enabling unique scientific applications, such as astrometric studies of southern massive star-forming regions \citep{2024IAUS..380..106H}. Another important scientific application is to increase the number of astrometric measurements beyond the Galactic center using the EAVN, where only a limited number of such results are currently available \citep{2026AJ....171..364S,2023PASJ...75..208S}.

During the KaVA observations, the peak flux density of W3(OH) varied from 510\,$\pm$\,50 to 3900\,$\pm$\,400 Jy (Figure~\ref{fig:7}). This study presents the first successful VLBI maps of W3(OH) obtained during such large flux density variations. This variation is primarily attributed to maser features ID1 and ID2, which are separated by only $\sim$64 AU at a distance of 2.01 kpc (Figures~\ref{fig:8} and \ref{fig:8b}). A clear trend of decreasing linewidth with increasing logarithm of the peak flux density was observed for maser feature ID1 (Pearson $r=-0.86$ ($p=0.03$) and Spearman $\rho=-0.89$ ($p=0.02$); Figure~\ref{fig:10}), whereas a similar relationship was not evident in maser feature ID2. The overall spatial distribution of maser spots in the W3(OH) region remained largely unchanged during the observations, except that maser features ID1 and ID2 exhibited significant flux variations (Figure~\ref{fig:9}). The observed flux variation in ID1 is consistent with an increase in the effective optical depth under unsaturated maser amplification, whereas maser feature ID2 may have responded differently to the same change in amplification conditions owing to differences in its saturation state and/or internal velocity structure.

\begin{ack}
We thank the anonymous referee for the valuable and constructive comments, which have significantly improved the quality of this manuscript. We would like to thank the members of the East Asian VLBI Network and the Korea–Japan Correlation Center (KJCC) for their support in observations, correlation, and data reduction.
\end{ack}




\bibliographystyle{apj}
\bibliography{reference}

\appendix 

\onecolumn
\section*{Supplementary materials}
\label{appendix:1}

\begin{figure*}[tbhp] 
 \begin{center} 
     \includegraphics[scale=1.0]{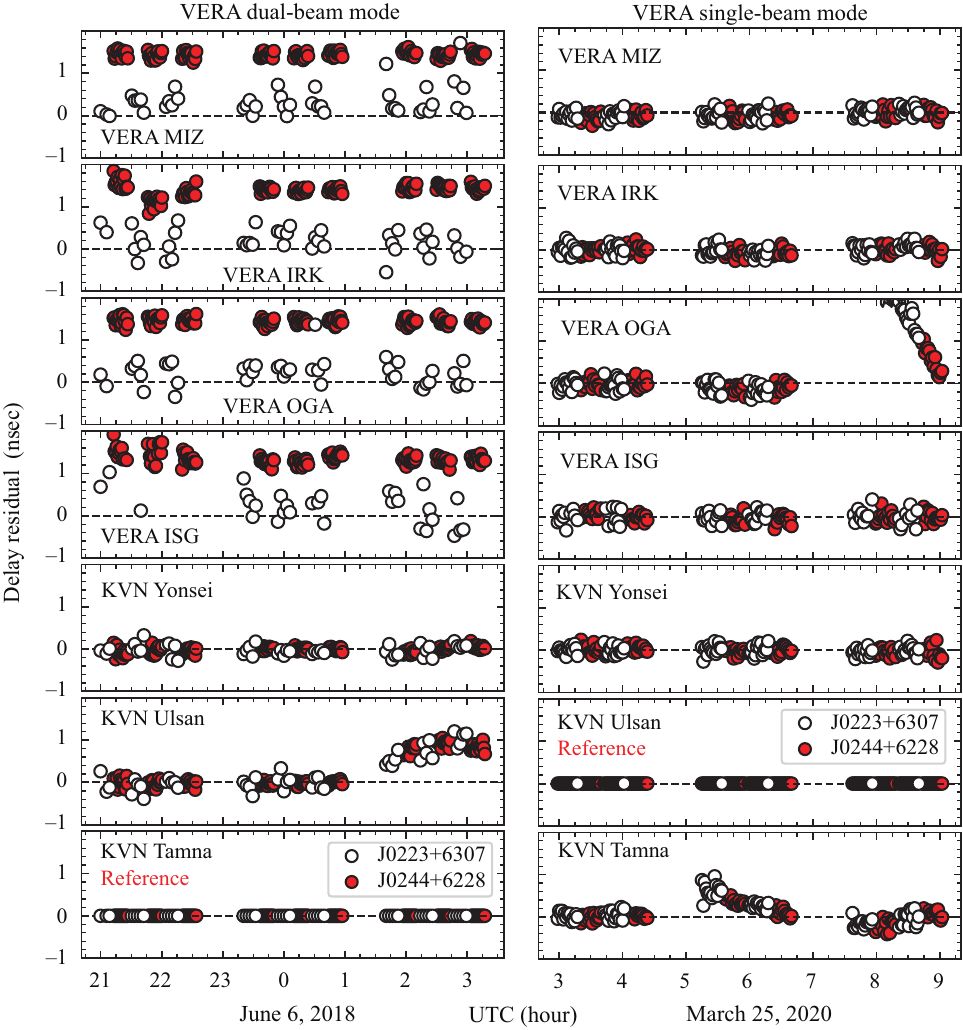} 
\end{center} 
\caption{Systematic delay offsets associated with the VERA dual-beam mode. (Left) Fringe solutions (i.e., delay residuals) for the phase-reference sources J0223+6307 (white) and J0244+6228 (red) at the VERA Mizusawa, Iriki, Ogasawara, and Ishigaki-jima stations, and the KVN Yonsei, Ulsan, and Tamna stations, from top to bottom. The reference antenna is KVN Tamna. The data were obtained on 2018 June 6, when the VERA dual-beam mode was used. (Right) Same as the left panels, but for observations obtained with the VERA single-beam mode on 2020 March 25. The reference antenna is KVN Ulsan. {Alt text: Delay residuals as a function of time for multiple antennas, comparing the VERA dual-beam and single-beam observing modes.}} 
\label{fig:3} 
\end{figure*}

\begin{figure*}[tbhp] 
 \begin{center} 
     \includegraphics[scale=1.0]{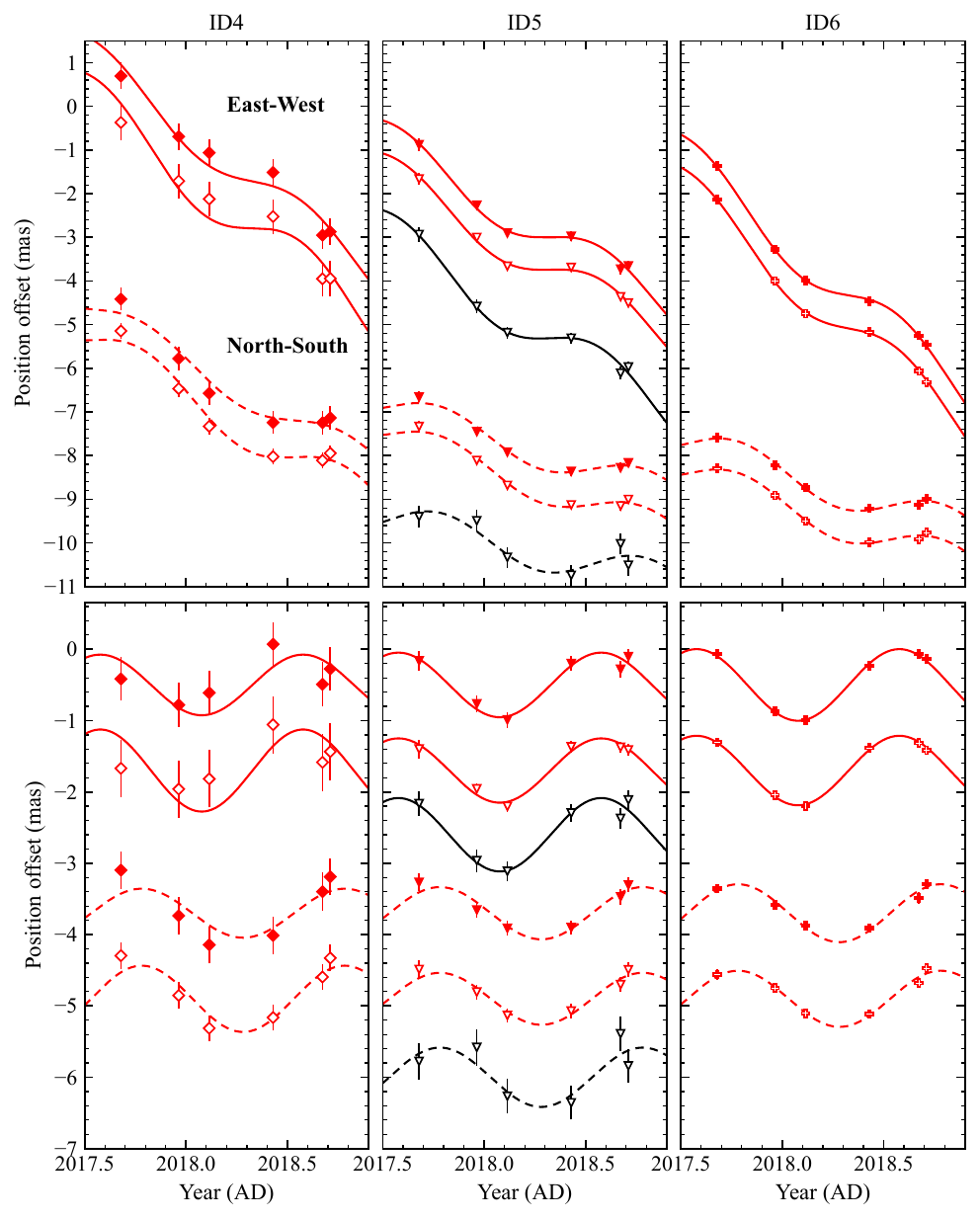} 
\end{center} 
\caption{Same as Figure \ref{fig:4}, but for maser features ID4, ID5, and ID6. {Alt text: Panels showing position offsets as a function of time for maser features ID4, ID5, and ID6. The top row shows the full astrometric fits, while the bottom row shows the parallax component after removing the fitted proper motions.}}
\label{fig:4b} 
\end{figure*}

\begin{figure*}[tbhp] 
 \begin{center} 
     \includegraphics[scale=1.0]{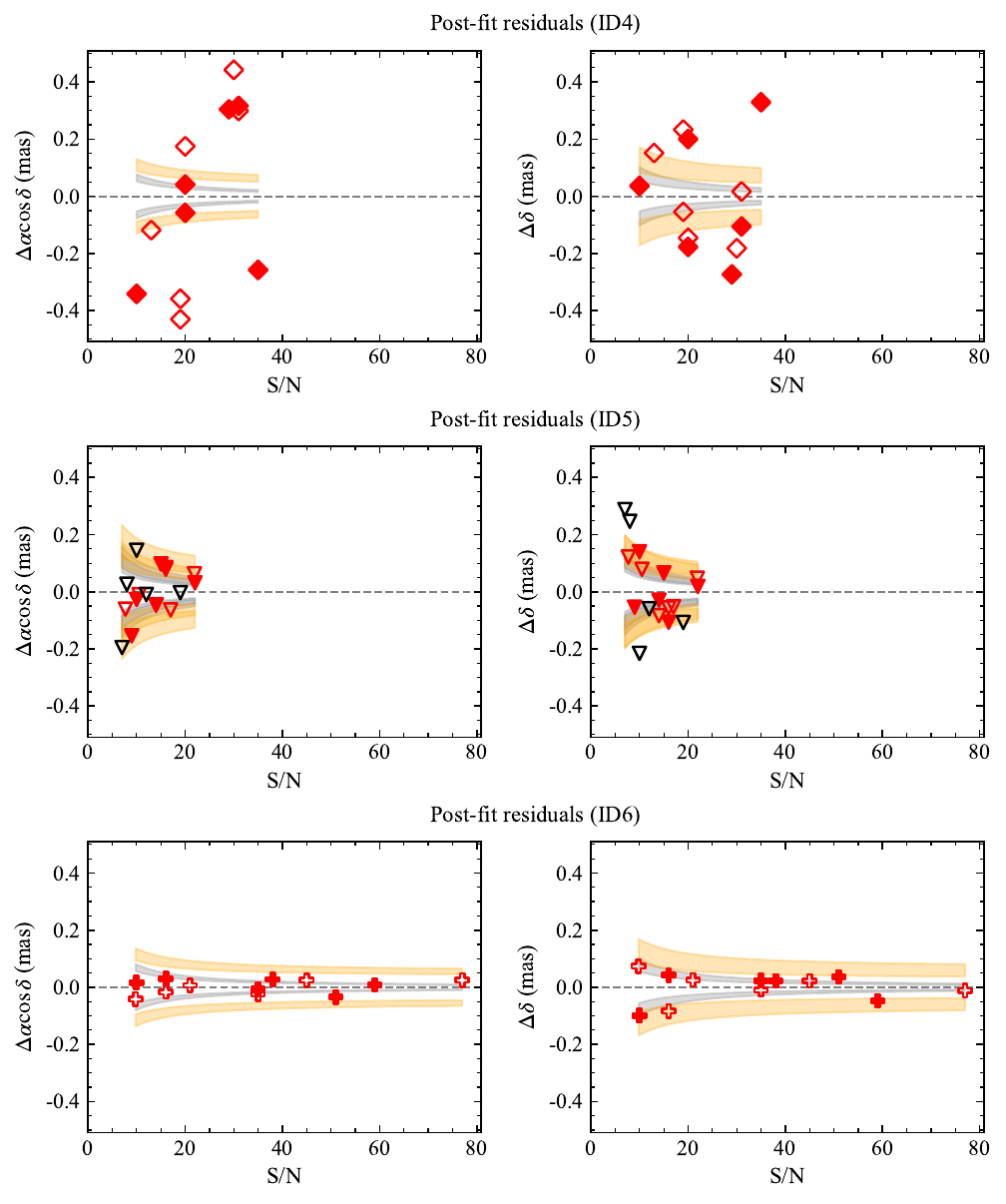} 
\end{center} 
\caption{Same as Figure \ref{fig:6}, but for maser features ID4 (top row; see Table \ref{table:6}), ID5 (middle row), and ID6 (bottom row). {Alt text: Panels showing the post-fit residuals in right ascension (left column) and declination (right column) for maser features ID4 (top row), ID5 (middle row), and ID6 (bottom row) as a function of the signal-to-noise ratio (S/N) of the phase-referenced images.}} 
\label{fig:6b} 
\end{figure*}

\begin{figure}[tbhp] 
 \begin{center} 
     \includegraphics[scale=1.0]{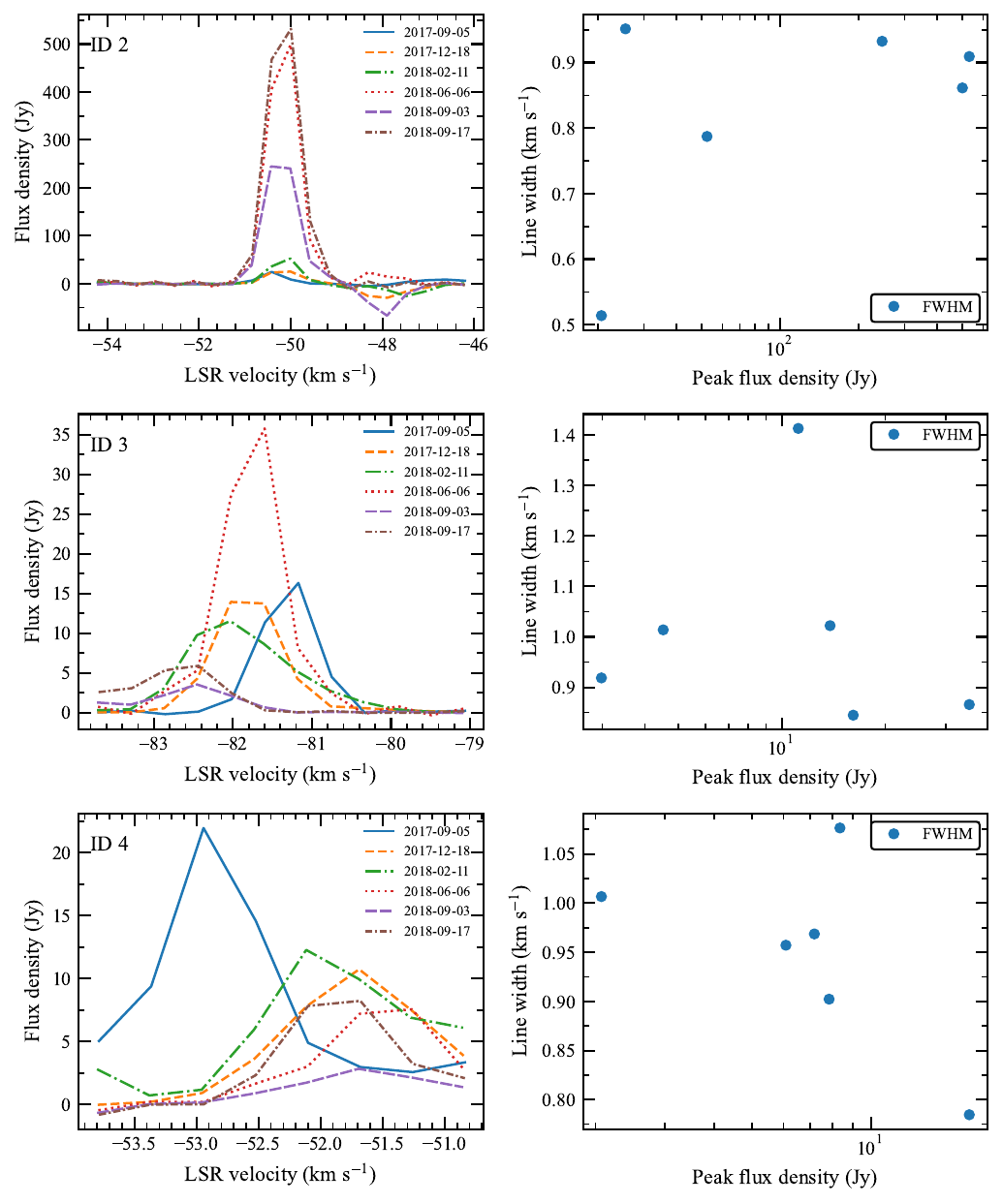} 
\end{center} 
\caption{{\bf(Left)} Same as Figure \ref{fig:8}, but for maser features ID2 (top), ID3 (middle), and ID4 (bottom). {\bf(Right)} Same as Figure \ref{fig:10}, but for the spectra shown in the left panels. The Pearson correlation coefficients are $r = 0.53$ ($p = 0.28$), $r = -0.11$ ($p = 0.83$), and $r = -0.60$ ($p = 0.21$), while the corresponding Spearman rank correlation coefficients are $\rho = 0.26$ ($p = 0.62$), $\rho = -0.43$ ($p = 0.40$), and $\rho = -0.37$ ($p = 0.47$) for maser features ID2, ID3, and ID4, respectively. {Alt text: Time-series spectra and FWHM line width as a function of peak maser flux density for maser features ID2, ID3, and ID4.}}
\label{fig:8b} 
\end{figure}

\begin{figure}[tbhp] 
 \begin{center} 
     \includegraphics[scale=1.0]{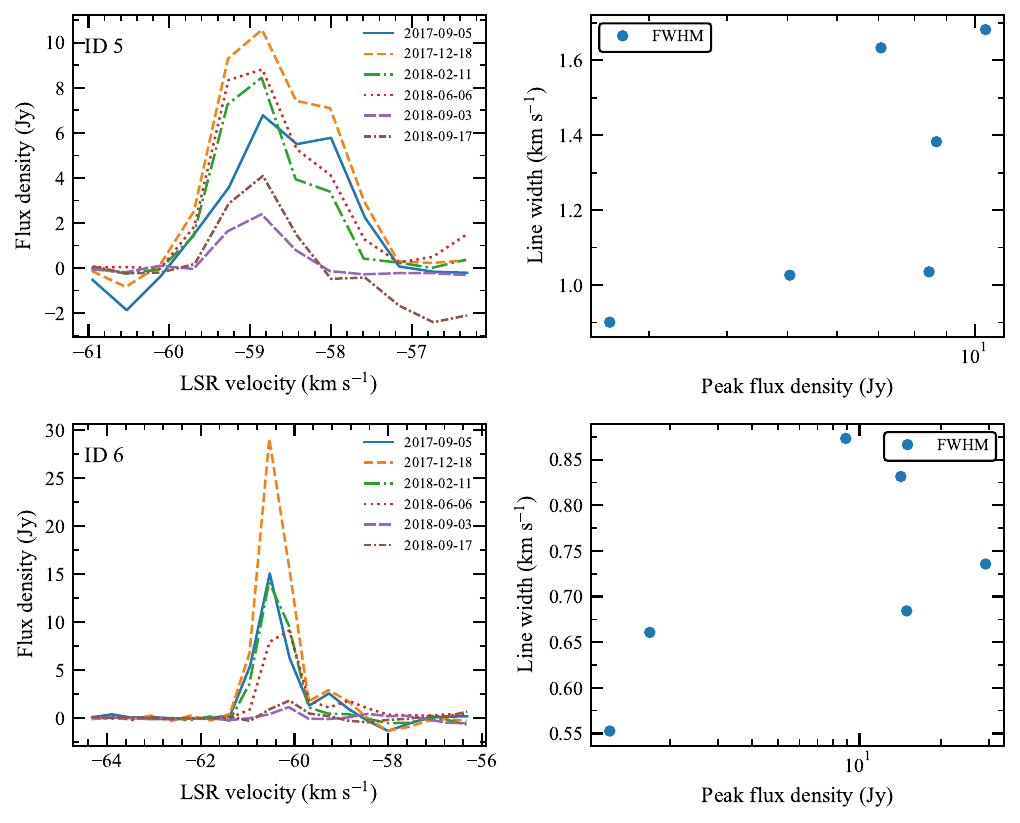} 
\end{center} 
\caption{Same as Figure~\ref{fig:8b}, but for maser features ID5 (top) and ID6 (bottom). The Pearson correlation coefficients are $r = 0.70$ ($p = 0.12$) and $r = 0.65$ ($p = 0.16$), while the corresponding Spearman rank correlation coefficients are $\rho = 0.83$ ($p = 0.04$) and $\rho = 0.49$ ($p = 0.33$) for maser features ID5 and ID6, respectively.
{Alt text: Time-series spectra and FWHM line width as a function of peak maser flux density for maser features ID5 and ID6.}}
\label{fig:8c} 
\end{figure}

We present supplementary information for this paper. Figure \ref{fig:3} provides supplementary material on the residual instrumental delays associated with the VERA dual-beam mode discussed in the main text. Figure \ref{fig:4b} presents the parallax and proper-motion fitting results for maser features ID4, ID5, and ID6. Figure \ref{fig:6b} shows the post-fit residuals of the parallax and proper-motion fits for maser features ID4, ID5, and ID6. Figures \ref{fig:8b} and \ref{fig:8c} show the temporal variations of the spectra and the relationship between line width and peak flux density for maser features ID2 through ID6. Figure \ref{fig:8b} presents the results for features ID2–ID4, while Figure \ref{fig:8c} presents those for features ID5 and ID6.

\end{document}